\pageno=1                                      
%
%
%
%
\font\ninerm=cmr9
\font\eightrm=cmr8
\font\sixrm=cmr6
\font\ninei=cmmi9
\font\eighti=cmmi8
\font\sixi=cmmi6
\skewchar\ninei='177 \skewchar\eighti='177 \skewchar\sixi='177
\font\ninesy=cmsy9
\font\eightsy=cmsy8
\font\sixsy=cmsy6
\skewchar\ninesy='60 \skewchar\eightsy='60 \skewchar\sixsy='60

\font\ninebf=cmbx9
\font\eightbf=cmbx8
\font\sixbf=cmbx6
\font\ninett=cmtt9
\font\eighttt=cmtt8
\hyphenchar\tentt=-1 
\hyphenchar\ninett=-1
\hyphenchar\eighttt=-1
\font\ninesl=cmsl9
\font\eightsl=cmsl8
\font\nineit=cmti9
\font\eightit=cmti8
\newskip\ttglue
\def\tenpoint{\def\rm{\fam0\tenrm}%
  \textfont0=\tenrm \scriptfont0=\sevenrm \scriptscriptfont0=\fiverm
  \textfont1=\teni \scriptfont1=\seveni \scriptscriptfont1=\fivei
  \textfont2=\tensy \scriptfont2=\sevensy \scriptscriptfont2=\fivesy
  \textfont3=\tenex \scriptfont3=\tenex \scriptscriptfont3=\tenex
  \def\it{\fam\itfam\tenit}%
  \textfont\itfam=\tenit
  \def\sl{\fam\slfam\tensl}%
  \textfont\slfam=\tensl
  \def\bf{\fam\bffam\tenbf}%
  \textfont\bffam=\tenbf \scriptfont\bffam=\sevenbf
   \scriptscriptfont\bffam=\fivebf
  \def\tt{\fam\ttfam\tentt}%
  \textfont\ttfam=\tentt
  \tt \ttglue=.5em plus.25em minus.15em
  \normalbaselineskip=12pt
  \let\sc=\eightrm
  \let\big=\tenbig
  \setbox\strutbox=\hbox{\vrule height8.5pt depth3.5pt width0pt}%
  \normalbaselines\rm}
\def\ninepoint{\def\rm{\fam0\ninerm}%
  \textfont0=\ninerm \scriptfont0=\sixrm \scriptscriptfont0=\fiverm
  \textfont1=\ninei \scriptfont1=\sixi \scriptscriptfont1=\fivei
  \textfont2=\ninesy \scriptfont2=\sixsy \scriptscriptfont2=\fivesy
  \textfont3=\tenex \scriptfont3=\tenex \scriptscriptfont3=\tenex
  \def\it{\fam\itfam\nineit}%
  \textfont\itfam=\nineit
  \def\sl{\fam\slfam\ninesl}%
  \textfont\slfam=\ninesl
  \def\bf{\fam\bffam\ninebf}%
  \textfont\bffam=\ninebf \scriptfont\bffam=\sixbf
   \scriptscriptfont\bffam=\fivebf
  \def\tt{\fam\ttfam\ninett}%
  \textfont\ttfam=\ninett
  \tt \ttglue=.5em plus.25em minus.15em
  \normalbaselineskip=10pt 
  \let\sc=\sevenrm
  \let\big=\ninebig
  \setbox\strutbox=\hbox{\vrule height8pt depth3pt width0pt}%
  \normalbaselines\rm}
\def\eightpoint{\def\rm{\fam0\eightrm}%
  \textfont0=\eightrm \scriptfont0=\sixrm \scriptscriptfont0=\fiverm
  \textfont1=\eighti \scriptfont1=\sixi \scriptscriptfont1=\fivei
  \textfont2=\eightsy \scriptfont2=\sixsy \scriptscriptfont2=\fivesy
  \textfont3=\tenex \scriptfont3=\tenex \scriptscriptfont3=\tenex
  \def\it{\fam\itfam\eightit}%
  \textfont\itfam=\eightit
  \def\sl{\fam\slfam\eightsl}%
  \textfont\slfam=\eightsl
  \def\bf{\fam\bffam\eightbf}%
  \textfont\bffam=\eightbf \scriptfont\bffam=\sixbf
   \scriptscriptfont\bffam=\fivebf
  \def\tt{\fam\ttfam\eighttt}%
  \textfont\ttfam=\eighttt
  \tt \ttglue=.5em plus.25em minus.15em
  \normalbaselineskip=9pt
  \let\sc=\sixrm
  \let\big=\eightbig
  \setbox\strutbox=\hbox{\vrule height7pt depth2pt width0pt}%
  \normalbaselines\rm}
%
\def\headtype{\ninepoint}                 
\def\abstracttype{\ninepoint}             
\def\captiontype{\ninepoint}              
\def\footnotetype{\ninepoint}             
\def\refit{\it}                           
\font\chaptitle=cmr10 at 11pt             
\rm                                       

%
%
\parindent=0.25in                         
\parskip=0pt                              
\baselineskip=12pt                        
\hsize=4.25truein                         
\vsize=7.445truein                        
\hoffset=1in                              
\voffset=-0.5in                           

\newskip\sectionskipamount                
\newskip\aftermainskipamount              
\newskip\subsecskipamount                 
\newskip\firstpageskipamount              
\newskip\capskipamount                    
\newskip\ackskipamount                    
\sectionskipamount=0.2in plus 0.09in
\aftermainskipamount=6pt plus 6pt         
\subsecskipamount=0.1in plus 0.04in
\firstpageskipamount=3pc
\capskipamount=0.1in
\ackskipamount=0.15in
\def\sectionskip{\vskip\sectionskipamount}
\def\aftermainskip{\vskip\aftermainskipamount}
\def\subsecskip{\vskip\subsecskipamount} 
\def\firstpageskip{\vskip\firstpageskipamount}

%
%
\nopagenumbers                            
\newcount\firstpageno                     
\firstpageno=\pageno                      
\newcount\chapno                          

\def\rightheadline{\headtype\phantom{\folio}\hfil\runningtitletext\hfil\folio}
\def\leftheadline{\headtype\folio\hfil\runningauthortext\hfil\phantom{\folio}}
\headline={\ifnum\pageno=\firstpageno\hfil
           \else
              \ifdim\ht\topins=\vsize           
                 \ifdim\dp\topins=1sp \hfil     
                 \else
                     \ifodd\pageno\rightheadline\else\leftheadline\fi
                 \fi
              \else
                 \ifodd\pageno\rightheadline\else\leftheadline\fi
              \fi
           \fi}

\def\bottomnumber{\hss\tenrm[\folio]\hss}
\footline={\ifnum\pageno=\firstpageno\bottomnumber\else\hfil\fi}

%
%
%
%
\outer\def\mainsection#1
    {\vskip 0pt plus\smallskipamount\sectionskip
     \message{#1}\vbox{\noindent{\bf#1}}\nobreak\aftermainskip\noindent}
 
\outer\def\subsection#1
    {\vskip 0pt plus\smallskipamount\subsecskip
     \message{#1}\vbox{\noindent{\bf#1}}\nobreak\smallskip\nobreak\noindent}
 

\def\title#1{{\chaptitle\leftline{#1}}}
\def\name#1{\leftline{#1}}
\def\affiliation#1{\leftline{\it #1}}
\def\abstract#1{{\abstracttype \noindent #1 \smallskip\vskip .1in}}
\def\ref{\noindent \parshape2 0truein 4.25truein 0.25truein 4truein}
\def\caption{\noindent \captiontype
             \parshape=2 0truein 4.25truein .125truein 4.125truein}

\def\footnote#1{\edef\fspafac{\spacefactor\the\spacefactor}#1\fspafac
      \insert\footins\bgroup\footnotetype
      \interlinepenalty100 \let\par=\endgraf
        \leftskip=0pt \rightskip=0pt
        \splittopskip=10pt plus 1pt minus 1pt \floatingpenalty=20000
        \textindent{#1}\bgroup\strut\aftergroup\strut\egroup\let\next}
\skip\footins=12pt plus 2pt minus 4pt 
\dimen\footins=30pc 

%
%

\def\@{\spacefactor 1000}

\def\,{\pcomma} 
\def\pcomma{\relax\ifmmode\mskip\thinmuskip\else\thinspace\fi}

\def\oversim#1#2{\lower0.5ex\vbox{\baselineskip=0pt\lineskip=0.2ex
     \ialign{$\mathsurround=0pt #1\hfil##\hfil$\crcr#2\crcr\sim\crcr}}}
\def\simgt{\mathrel{\mathpalette\oversim>}}
\def\simlt{\mathrel{\mathpalette\oversim<}}
%
%
\voffset 1cm
%
%
%

\def\arcsec{\hbox{$^{\prime\prime}$}}

\def\farcs{\hbox{$.\!\!^{\prime\prime}$}}
\def\micron{\hbox{$\mu$m}}
%
%
\def\msun{\hbox{$M_{\odot}$}}
\def\rsun{\hbox{$R_{\odot}$}}  
\def\runningtitletext{YOUNG BINARY STARS AND ASSOCIATED DISKS}
\def\runningauthortext{Robert D. Mathieu et al.}

\null
\firstpageskip

{\baselineskip=14pt
\title{YOUNG BINARY STARS AND ASSOCIATED DISKS}
}

\vskip .3truein
\name{ROBERT D. MATHIEU}
\affiliation{University of Wisconsin - Madison}
\vskip .2truein
\name{ANDREA M. GHEZ}
\affiliation{University of California, Los Angeles}
\vskip .2truein
\name{ERIC L. N. JENSEN}
\affiliation{Swarthmore College}
\vskip .2truein
\name{MICHAL SIMON}
\affiliation{State University of New York, Stony Brook}
\vskip .3truein

\abstract{The typical product of the star formation process is a
multiple star system, most commonly a binary star. 
Binaries have provided the first dynamical measures of the masses of
pre-main-sequence (PMS) 
stars. These measurements have established that T Tauri-like stars are
indeed of solar mass or less, and have provided preliminary support for the mass
calibrations of theoretical PMS evolutionary tracks.
Surprisingly, in some star-forming regions PMS binary frequencies have been 
found to be higher than 
among main-sequence solar-type stars. The binary frequency in the Taurus 
star-forming region is a factor two in excess of the field, although other regions
show no excess. Observations suggest that the difference in PMS and main-sequence
binary frequencies is
not an evolutionary effect,
and thus recent attention has focussed on correlations between binary frequency
and initial conditions (e.g., stellar density or cloud temperatures).
Accretion disks are common among young binary stars. Binaries with separations
between 1 AU and 100 AU have substantially
less submillimeter emission than closer or wider binaries, suggesting that such
binaries have dynamically truncated their associated disks. Direct evidence of
dynamical clearing has been seen in several binaries, most notably GG Tau.
Remarkably, PMS binaries of all separations show evidence of long-lived 
circumstellar disks and continued accretion at stellar surfaces. This strongly
suggests that the circumstellar disks are replenished from circumbinary disks
or envelopes, perhaps through recently hypothesized accretion streams across
dynamically cleared gaps.
The frequent presence of either circumstellar or circumbinary disks suggests
that planet formation can occur in binary environments. That planets may
form around stars in wide binaries is already established by their discovery.
Circumbinary disk masses around very short period binaries are ample to form
planetary systems such as our own. The nature of planetary systems among the
most frequent binaries, with separations between 10 AU and 100 AU, is less
clear given the observed reduction in disk mass.
However, even these systems have disks with masses adequate for
the formation of terrestrial-like planets.
}


\mainsection{I.~~INTRODUCTION}
During the last decade we have gone from {\it suspecting} that most pre-main-sequence
(PMS) stars are binary stars to {\it knowing} that most PMS stars are
binary stars. Advances in high-angular-resolution infrared
imaging technology have enabled 
large surveys for binaries in a variety of
star-forming regions. The consistent result is that binary stars are abundant
among young stars, indeed perhaps remarkably so.

The implications of this for star and planet formation research are
enormous.  From
an observational point of view we ignore the presence of companion stars at our
own peril. A companion star may 
substantially contaminate the light
attributed to the primary, thereby leading to incorrect luminosity, temperature
and extinction measurements, erroneous age and mass estimates, incorrect disk
models, and so on.  From a theoretical point of view, we
must recognize that binary star formation is the primary branch of the 
star formation process. While our presence in orbit around a single star will
always drive interest in single-star formation, a general understanding of
star formation must focus on multiple star formation. Similarly, the typical
environment for planet formation may be a binary star. 
The ubiquity of binary
systems suggests that the question of planet formation
in binary systems is critical for setting the overall
frequency of planets.

In this contribution we will focus on those observations of PMS
binaries as they relate to (1) probes of evolutionary models,
(2) binary star populations, and (3) protoplanetary disks and
the potential for planet formation. 
Typically, consideration will be restricted to binaries with
primaries having masses less than 3 \msun\ (with primaries of solar
mass or less predominating) and ages of less than 10 million
years. Observations of such PMS binaries have been comprehensively
reviewed by Mathieu (1994; see also review papers in Milone and Mermilliod 
1996), and thus we will concentrate our attention
on observations made since then. Many theoretical issues of binary formation are
addressed in Bodenheimer et al./ (this volume).
\mainsection{{I}{I}.~~PRE-MAIN-SEQUENCE STELLAR PROPERTIES}
Typically, the masses and ages of young stars are
inferred from comparisons with PMS evolutionary tracks. The mass
and age calibrations of such tracks are largely untested against observations.
Young binary stars provide powerful tools for assessing the validity of
these models.  In particular, their orbital motions yield {\it direct}
measures of their stellar masses, while under assumptions of coeval
formation the derived ages of binary components test relative age
calibrations of the tracks.

High-angular-resolution observations from both speckle 
interferometry and the Hubble Space 
Telescope Fine Guidance Sensors have begun to contribute
significantly to the determination of astrometric orbits (Ghez et al.\ 1995;
Thi\'ebaut et al.\ 1995; Simon et al.\ 1996). These studies clearly show relative motion of the two stars
in many PMS binaries, and show 
curvature in the relative motions of several. 
These motions are larger than the velocity
dispersions in star-forming regions (SFRs) and thus generally argue for an orbital
origin.

Ghez et al.\ (1995) treated the observed relative motions statistically to 
obtain a typical system mass among a set of astrometric binaries.  Their
results are shown in Fig.\ 1. The trend in the data of increasing 
relative velocity with decreasing separation also strongly suggests orbital 
motion. The comparison curves represent the
expected values of relative velocity as a function of binary separation for a
set of binary stars observed at random orientations. While any given datum reflects 
only the state of a binary at a certain moment in its orbit, and thus cannot
reliably provide binary mass, the median mass of 1.7 \msun\ found from the 
ensemble should be a valid estimate of the typical binary total mass. This 
value empirically supports the theoretical conclusion that most PMS stars have 
solar masses or less.

Eclipsing binaries are particularly powerful routes to the measurement of {\it
both} stellar masses and radii. Until recently only EK Cep had been studied in
detail (Popper 1987; Claret et al.\ 1995; Mart\'{\i}n and Rebolo 1993). Unfortunately
the PMS secondary of this system is very near the ZAMS and thus places little
constraint on evolutionary tracks. A second eclipsing binary, TY~CrA, recently
has been intensively studied by two groups (Casey et al.\ 1998, 
Corporon et al.\ 1996, and references therein). 
This system consists of a Herbig Be star of 3.16 \msun\ on the main sequence 
and a cool (4900 K) 1.64 \msun\ secondary star. The 2.1 \rsun\ radius of the 
secondary, along with its association with the R CrA dark cloud and lithium 
absorption, identify it as PMS\null. Comparison
with evolutionary models places it at the base of its Hayashi track, with an age
of 3 Myr (Fig.\ 2). Casey et al.\ tested three sets of evolutionary tracks---
Swenson et al.\ (1994; Fig.\ 2), Claret (1995), and D'Antona and Mazzitelli
(1994)---against the TY~CrA secondary. In all three cases solar-composition
1.64 \msun\ tracks are consistent with the observed physical parameters. 
Thus the secondary star represents the first quantitative 
dynamical test of PMS evolutionary
tracks, which they pass without contradiction. 
Unfortunately, the accuracies of the derived physical parameters were 
not adequate
to distinguish critically {\it among} evolutionary tracks. Discovery and 
study of additional PMS eclipsing binaries are sorely needed. 

Stellar masses can also be measured via orbital motions of
disk gas.  The circumstellar disks of the single stars DM Tau and GM Aur, and
the circumbinary disks of GG Tau and UY Aur have been used in this
way (Dutrey et al.\ 1994; Duvert et al.\ 1998; Guilloteau and Dutrey 1998;
Dutrey et al.\ 1996; Guilloteau et al. 1999).   
In well-positioned systems, the relative precision
of this technique can be less than 10\%, so that uncertainty in the
distance to a given system 
is the limiting factor in the determination of its mass. 

The {\it relative} mass calibrations of evolutionary models can be tested with
careful analyses of PMS double-lined binaries, which provide very accurate
dynamical mass ratios. Adopting the observations and analysis procedures
of Lee (1992), Figueiredo (1997) has used the PMS binary 162814$-$2427 to test
sets of PMS evolutionary models with different input physics; Lee also
studied four other double-lined systems. Both Lee and Figueiredo find that the 
relative mass calibrations of evolutionary tracks are consistent with the
observed mass ratios of the 
binaries, presuming coeval formation of the component 
stars. However, Figueiredo notes that
the observational uncertainties are considerably higher than
the theoretical ones. {\it More generally, it should be appreciated that even given
accurate mass determinations meaningful comparison with theory is severely 
limited by uncertainties in the effective temperatures and luminosities of the 
weighed stars.} (See Fig.\ 3 for an example of one aspect of the temperature
difficulty.)

Recently, Prato (1998) has extended this observational test into the
near-infrared. Infrared observations permit the detection of cooler
companions than does the optical, which has the advantages of 1) converting
single-lined systems into double-lined systems and thus expanding the sample
and 2) providing double-lined systems with large mass ratios which give more
leverage in testing the models. Prato has obtained H-band high-resolution
spectroscopy of 
the previously single-lined system NTT 155913-2233, and detected an M5 companion
to the K5 primary. The derived mass ratio is about 2, the largest mass ratio
yet measured among PMS binaries. Interestingly only the Swenson et al.\ (1994)
models are roughly (within 1 $\sigma$) consistent with the binary system. The
components are neither coeval nor consistent with the mass limits using
the D'Antona and Mazzitelli tracks.

Similarly, White et al.\ (1999) conducted a test of evolutionary 
models by requiring that the components of the quadruple GG Tau be the same age.
This system is particularly useful as its components span a wide range in 
mass and extend across the stellar/substellar boundary, a region
where both the evolutionary models and the PMS temperature
scale are very uncertain.  Of the evolutionary models tested,
they find the Baraffe et al.\ (1998) models yield the most consistent
ages using a temperature scale intermediate between that of giants and dwarfs 
for PMS M stars (Fig.\ 3).  
With this model, the coldest component of the GG Tau
system, GG Tau Bb, is substellar with a mass of 50 $M_{Jup}$. 

At present we have no empirical measure of the absolute ages of PMS stars,
and thus age determination relies on the application of evolutionary models to
the temperatures and luminosities of the stars. As stressed by Simon et al.\
(1993), applying this procedure to the combined light of binaries must lead to
biased results, primarily toward younger ages due to enhanced luminosity. 
Typical errors are of order a factor 2, although errors as large as an order
of magnitude are possible (Brandner and Zinnecker 1997, Ghez et al.\ 1997b).

Assuming a given evolutionary model, the inferred
relative ages of the components of binaries provide an empirical test 
of coevality, which is useful for distinguishing between different
modes of binary star formation.  Hartigan
et al.\ (1994) determined the ages of components of wide binaries, and found that
in roughly one-third of the cases the secondary was significantly younger than
the primary. Brandner and Zinnecker (1997) did not find any such age
differences in a sample of somewhat closer binaries. Ghez et al.\ (1997) noted
that many of the binary components in the Hartigan et al.\ sample were themselves
binaries, leading to biased age estimates. When they considered the subset of
the Hartigan et al.\ sample not known to be triples, all of the binary components
were coeval to within the uncertainties.

\mainsection{{I}{I}{I}.~~YOUNG BINARY POPULATIONS}
The early multiplicity surveys of the Taurus and Ophiuchus dark cloud complexes 
revealed an apparent difference in the binary star fractions of 
PMS and main-sequence stars  (Ghez et al.\ 1993; 
Leinert et al.\ 1993; Simon et al.\ 1995).  This difference was found to be
particularly pronounced within the separation range of 1 AU to 150 
AU, where the young stars were found to be twice 
as likely to be members of a binary 
system as the older stars observed by Duquennoy and Mayor (1991, DM91).  
Interpretations of this unexpected finding have included 
(1) observational selection effects, especially differences in detection limits
between the surveys and differences in 
sample populations (e.g., relative fractions 
of classical T Tauris stars (CTTS) and
weak-lined T Tauri stars (WTTS)),  (2) formation differences among SFRs 
(e.g., Reipurth and Zinnecker 1993), (3) varying period distributions 
(Leinert et al.), or (4) the disruption of primordial multiple
star systems over time (Ghez et al.\ 1993). Alternatively, Mathieu (1996)
questioned whether the excesses were statistically significant.
The variety of interpretations led to
many subsequent surveys in order to increase both the sample sizes and
the numbers of SFRs studied (see Table 1). These surveys have shed light on
many of the questions raised.

\subsection{A.~~Is the excess of young companion stars real?}

The question of whether or not the excess frequency of PMS binaries
is real can be divided
into two parts.  First, is the high fraction of binary stars
in Taurus and Ophiuchus statistically significant?  
Second, is this really different
from the binary fraction in the solar neighborhood? The issue of the latter
question is incompleteness, and raises the related question of how incomplete
surveys should be compared.

Addressing the first question requires observation of larger samples
of stars.
K\"ohler and Leinert (1998) roughly doubled the sample
studied in the Taurus SFR by surveying 75 new PMS
stars discovered by $ROSAT$, most of which are classified as 
WTTS.\footnote{$^\dagger$}{It should be noted that distances, and hence
luminosities and ages, of the {\it ROSAT\/} selected stars
are controversial; Brice\~no et al.\ (1997) argued that the majority
are actually foreground ZAMS stars.  Neuh\"auser and Brandner (1998)
studied the small sample of these stars that are sufficiently bright
to have {\it HIPPARCOS\/} parallaxes, and found them to be younger
than $1.6 \times 10^7$ years, but in front of their presumed birth
place, possibly due to ejection.
Nonetheless, K\"ohler and Leinert point out that if their sample does include
foreground stars with a binary frequency like that of the nearby
solarlike stars, then the binary frequency 
of the WTTS remaining in the Taurus SFR must be
even higher than the observed value.}
Their survey is comparable in sensitivity to the previous infrared speckle
work done in Taurus (flux ratio $\Delta {\rm K}\sim 3$ mag) and results in
a binary frequency (BF, number of multiples divided by number of systems, 
hereinafter and in Table 1) consistent with the earlier surveys,
reinforcing the conclusion that Taurus does indeed have a remarkably high BF. 

The expanded Taurus sample has also addressed the concern that
CTTS and WTTS might have differing binary frequencies. This would
constitute a serious selection effect as the early surveys comprised 
primarily CTTS (which dominated the available catalogs such as
Herbig and Bell 1988), while in fact the X-ray selected WTTS appear to outnumber
the CTTS (e.g., Walter et al.\ 1994, Neuh\"auser 1997, Walter et al.\, this
volume).
However, K\"ohler and Leinert (1998) find no significant difference between
the CTTS and WTTS in either their BF or in their distribution of separations.
 
Although the measured binary frequency for PMS stars in Taurus is
a factor of two larger than that measured for stars in the solar neighborhood,
this does not necessarily mean that the parent populations
are different.  The discrepancy between the two binary frequencies could 
be due to a difference in sensitivity to low-mass companions.  In 
particular, it is possible that PMS star surveys are detecting very low-mass 
stars that are relatively more luminous when they are young
(e.g., Burrows et al.\ 1993, Malkov et al.\ 1998).  
Surveys of main-sequence stars generally
have well-defined mass-ratio ($q$) sensitivity limits.  In the case of the DM91
survey of solar-mass stars, 
the spectroscopic portion of the survey, which cover periods less
than $\sim$ 10,000 days or equivalently semi-major axes less than $\sim$ 10 AU,
is complete down to $q > 0.1$.   
DM91's sensitivity to
longer periods or more widely separated systems, identified by direct imaging,
is limited to $q > 0.3 $.   
In contrast, the limits of
PMS star surveys, which have generally been carried out at 
a single wavelength (typically 2.2 \micron), 
are harder to characterize in terms of mass and 
thus are generally described in terms of limiting flux ratios or flux 
densities.  Estimates of the secondary
star masses from these single wavelength measurements involve a number
of assumptions.  In particular, one has to assume that the two stars
have the same age, the same line-of-sight extinction, and no infrared excess
(e.g., Meyer and Beckwith 1998).  Follow-up studies, which resolve the binary stars
at multiple wavelengths, show that most systems have $q > 0.3$
(Hartigan et al.\ 1994; Brandner and Zinnecker 1997; Ghez et al.\ 1997b),
suggesting that the high young-star BF is unlikely to arise from a multitude 
of binary star systems with $q < 0.3$.
Still, the lack of understanding of the true mass
limits of the young surveys and the limited depth of DM91's main-sequence survey
at comparable separations is a major weakness in the discussion of the 
relative BFs of the PMS and main-sequence stars.  

\subsection{B.~~Does the binary population evolve in time?}
 
If the PMS and main-sequence binary frequencies do differ, a possible
explanation is
the disruption of primordial multiple star systems over time (Ghez et al.\ 1993).
This has led several groups to pursue observations of binary frequencies in
open clusters with different ages.  Four clusters have now been studied 
intensively by Bouvier and his collaborators using adaptive optics
and by Patience and her collaborators using speckle imaging: 
$\alpha$
Per ($\sim 50$--$70$ Myr), Pleiades ($\sim 80$--$120$ Myr), Hyades ($\sim 600$ Myr)
and Praesepe ($\sim 600$ Myr).  

The K-band speckle imaging survey of the Hyades (Patience et al.\ 1998a),
$\alpha$ Per and Praesepe (Patience et al.\ 1998b) span a wide range
of spectral types (A0--K5). These studies have 
a uniform mass ratio limit over their separation range, which
fortuitously matches that of DM91; thus 
their uncorrected BF for F7--G9 stars is directly comparable to DM91's
sample. They find the cluster binary frequencies to be both 
statistically consistent with the solar neighborhood population and to be
significantly lower than the BF in Taurus.
The near-infrared adaptive-optics studies of G and K stars in the
Pleiades (Bouvier et al.\ 1997) and Praesepe (Bouvier et al., in preparation)
have less uniform mass ratio limits, ranging from 0.6 to  less than 0.1
over the separation range (0\farcs08--6\farcs9) studied.
Over the separation range reaching $q=0.3$ (0\farcs3--6\farcs9) and
limiting the BF to $q>0.3$ ratios results in a BF of 0.14
for the Pleiades, comparable to that reported by DM91 in this range (0.17);
Praesepe produces similar results.  The lack of change in BFs within this
age sequence of clusters indicates that evolution is not a strong effect, at
least after $\sim 50$--$70$ Myr. 
 
\subsection{C.~~Does the binary formation outcome vary among the SFRs?}

The apparent overabundance of young companions in Taurus with respect
to the solar neighborhood can
also be explained if Taurus-like SFRs are not the origin of most stars in  
our solar neighborhood.  This would require other SFRs 
to be less efficient at either forming or maintaining binary stars,
{\it and} to be the dominant contributors to the field population.
Consistent with this line of reasoning,
the majority of field stars have been suggested
to originate in dense stellar clusters in giant molecular clouds and
not in low-stellar-density regions such as the Taurus dark cloud complex
(e.g., Lada et al.\ 1991).  Furthermore, high-density SFRs might plausibly
have lower binary fractions, either by inhibiting binary
formation or by promoting their rapid destruction (e.g.\ through encounters 
(Kroupa 1995) or erosion of circumstellar disks (Hall 1997)).  

Orion, the closest giant molecular cloud,
is three times as distant as Taurus, and thus systems similar to
the closest binaries resolved in Taurus are not currently
resolved in Orion.  Nonetheless there is still a large
overlap in the separation/period range studied so far.  Petr et al.\ (1998) 
and Simon et al.\ (1999) have
studied the innermost region of the Orion Trapezium cluster using high-resolution
imaging
techniques at K\null. They find a BF similar to that of the nearby 
solar-like stars (Table 1), although the uncertainty is large owing to 
the small number in the sample.  Padgett et al.\ (1997) used V- and I-band 
$HST$ images of the Trapezium, and also of NGC 2024, 2068, and 2071, to 
investigate the BF\null. In both samples, they measured
a BF somewhat higher than that of the solar-like stars, but only at
a $\sim 1 \sigma$ level.  While these estimates are lower bounds because
they do not correct for incompleteness, taken together with Prosser
et al.'s (1994) results they suggest that the BF in the clusters of the
Orion OB association is similar to that of the  
nearby solar-like stars, at least in the range of separations surveyed to date.  

It is tempting to conclude that the high densities of the 
young clusters have led to lower binary frequencies than in Taurus, 
and more generally that
binary frequency is anti-correlated with stellar density. Several other
regions in Table 1 are consistent with this hypothesis (e.g., the dark
cloud complexes Chamaeleon, Corona Australis, and Lupus
(Ghez et al.\ 1997a); but see IC348 (Duchene et al. 1999a)), 
but the precisions of
the measured BFs are lower, as are the confidences of the 
consequent conclusions. In addition, the 
present samples do not clearly discriminate between stellar density and other
environmental factors. Thus Durisen and Sterzik (1994) have suggested that
cloud temperature may play a key role; in the existing samples the higher
stellar densities are found in OB associations which might arguably have higher
temperature environments than dark cloud complexes. Isolating the critical
factors in determining binary frequencies will be a challenging task.

Finally, none of these surveys covers the entire separation range of binary
populations. As such, differences in measured binary frequencies could be the 
result of differing separation distributions, perhaps also linked to initial
conditions. In this regard,
Brandner and K\"ohler (1998) report that the binary separation
distribution peaks at 90 AU
among the youngest subgroup in Sco OB, and at 215 AU toward an older subgroup
in the same association. As always, small sample sizes are a concern as is
the possibility of contamination by field stars at the larger separations.
The frequency of the shortest period binaries (e.g, $P < 100$ days) 
has also received attention. At present orbital solutions exist for nearly 30
spectroscopic binaries. In the most recent summary,
Mathieu (1996) reports 7 binaries
with $P < 100$ days among 91 PMS stars for a frequency of 8\% $\pm$ 3\%, the 
same as found for main-sequence solar-type stars. Even so, an excess 
comparable to those
found in the high-angular-resolution surveys is not excluded by
these data. Spectroscopic surveys of larger, carefully selected samples are
needed. High-precision long-time-scale surveys will be especially fruitful;
their sensitivity to long period systems will both increase the expected number
of detections, improving statistical significance, and provide important cases
for combined spectroscopic-astrometric orbital solutions.
\mainsection{{IV}.~~DISKS IN YOUNG BINARY STARS: STRUCTURE}

The frequency of binary companions is a critical datum with respect to
assessing the prospects for the formation of planets, in part
because of the impact of companions on protoplanetary disks.  
The observational case for disks in young binaries is well 
established (see Mathieu 1994).  Low-spatial-resolution observations
of CTTS binaries reveal many of the classic signatures of disk material and
accretion, such as excess emission at near-infrared through millimeter 
wavelengths, spectral veiling, Balmer and forbidden emission lines,
and polarization. Disk material
may be located around individual stellar companions (circumstellar disks) and/or 
entire binary star systems (circumbinary disks).
Theoretical
calculations of binary-disk interactions predict that companions will
truncate both circumstellar and circumbinary disks, with
circumstellar
disks having outer radii of 0.2--0.5 times the binary semimajor
axis $a$, and circumbinary disks having inner radii of 2$a$--3$a$,
with the exact values depending on eccentricity, mass ratio, and disk
viscosity (Artymowicz and Lubow 1994; see also Lubow and Artymowicz,
this volume).
Recent observations
have made significant progress towards delineating such structures of disks
in binary environments, and the potential
for planet formation in these disks.

\subsection{A.~~Disk masses}

Millimeter or submillimeter wavelength measurements of dust continuum
emission allow a measurement of the total disk mass present in a
system since at least part of the disk is optically thin at these
wavelengths.  The first systematic survey of millimeter emission from
a large number of young stars (Beckwith et al.\ 1990, Beckwith and Sargent
1993) suggested that
millimeter fluxes from close binaries might be lower than those from
wider binaries.  The subsequent discovery of many more young binaries
and further millimeter and submillimeter observations allowed more
detailed investigation of this question.  
Jensen et al.\ (1994, 1996a), Osterloh and Beckwith (1995), and
Nurnberger et al.\ (1998) found that millimeter fluxes (and by
extension, disk masses) are significantly lower among binaries with
separations of 1--100 AU than among wider binaries or single stars.
Binaries wider than 100 AU have a distribution
of millimeter fluxes indistinguishable from that of single stars.  
Finally, there is no evidence for a diminished disk mass around many PMS 
spectroscopic binaries (separations of less than 1 AU), including  GW Ori, 
UZ Tau E, AK Sco, DQ Tau and V4046 Sgr (Mathieu et al.\ 1995; 
Jensen et al.\ 1996a,b; Mathieu et al.\ 1997).

The amount of reduction in millimeter flux among the 1--100 AU
binaries is consistent with their circumstellar disks being truncated
at 0.2--0.5 times the binary separation, as predicted by theory.
However, the surface densities of these circumstellar disks are poorly
constrained. The low millimeter fluxes give typical disk-mass upper
limits of a few times $10^{-3}$ \msun, while the presence of {\it
IRAS\/} 12, 25, and 60 \micron\ emission from most of the binaries
requires the presence of at least tenuous circumstellar disks with
$M_{\rm disk} \simgt 10^{-5}$ \msun.
\footnote{$^\dagger$}{It should be noted that the absolute disk
masses are highly uncertain due to the poorly determined dust
opacities, gas-to-dust ratios, and disk surface-density profiles
(e.g., Beckwith and Sargent 1993).}

These observations suggest the rather intuitive picture that binaries
much wider or much closer than typical disk radii do not significantly
alter disk structure, while those binaries whose separations are
comparable to disk radii substantially modify the associated disks.
The breakpoint of roughly 100 AU is similar to the sizes of disks seen in
millimeter aperture-synthesis images (e.g., Koerner and Sargent 1995) and
optical/IR scattered-light images (McCaughrean et al., this volume).

Images of $\lambda=1.3$ mm continuum emission from the young quadruple
system UZ Tau empirically confirm this general picture on all three
scales (Fig.\ 4; Jensen et al.\ 1996b).
UZ Tau E is a spectroscopic binary with a projected
semimajor axis of $a\sin i = 0.1$ AU (Mathieu et al.\ 1996), and UZ
Tau W is a binary with a projected separation of 50 AU (Ghez et al.\
1993).  The two binaries are separated by roughly 500 AU\null.  UZ Tau
E has strong 1.3 mm emission that is resolved with a radius of $\sim$
170 AU and estimated mass of 0.06 \msun.  This circumbinary disk has a
size and mass similar to those seen around other young stars, and by
extension, similar to the early solar nebula.  It is not evidently
affected 
either by the presence of an embedded binary with separation
much smaller than its radius or by the presence of a companion (UZ Tau
W) at a separation much larger than its radius. In marked contrast, UZ
Tau W, with a separation comparable to a typical disk size, has
millimeter emission that is greatly reduced both in flux and in
spatial extent.  The unresolved millimeter emission must arise from
circumstellar disks around one or both of the stars in the 50 AU
binary.  It is noteworthy that, though the circumstellar disks are
reduced in mass and size, they are still present.  We note that
unresolved observations of the quadruple system would see a ``normal''
millimeter flux, i.e.\ one comparable to that from single stars,
whereas the distribution of disk material is in fact much more
complex.

\subsection{B.~~Circumbinary disks and disk clearing}
The millimeter surveys indicate that {\it massive} circumbinary disks are
rare in binaries wider than a few tens of AU\null.  As noted above, most
binaries with separations of 1--100 AU are undetected at millimeter
wavelengths, placing stringent limits on any circumbinary material.
In addition, Dutrey et al.\ (1996) made $\lambda=2.7$ mm interferometric
observations of 18 binaries in Taurus-Auriga and detected circumbinary
emission from only one, UY Aur. (Note that the circumbinary disk around GG Tau
had been previously discovered and was not included in this study.)  
In the widest binaries, stable
circumbinary material would lie at orbital distances of many hundreds
of AU and may be difficult to detect due to cold temperatures and low
surface densities. Nonetheless, substantial circumbinary disks would have been
detectable among binaries with separations of tens of AU.

The exceptions are notable, though. Both GG Tau (projected
separation 40 AU) and UY Aur (projected separation 120 AU) have
circumbinary material that is clearly seen in both millimeter
interferometric maps (Dutrey et al.\ 1994; Duvert et al.\ 1998; Guilloteau et 
al. 1999) and
near-infrared adaptive-optics images (Roddier et al.\ 1996; Close et
al.\ 1999; see also McCaughrean et al., this volume).  The material in
both of these circumbinary disks clearly shows Keplerian rotation.
In GG Tau, the circumbinary disk is resolved in both CO and continuum
emission (Fig.\ 5).  Detailed observations demonstrate that its
circumbinary disk has two components: a narrow ring with sharp edges including
about 80\% of the mass and an extended disk reaching as far as 800 AU.
The extended disk is cooler than the ring, consistent with heating by the star
and circumbinary disk.  The total circumbinary mass is 0.12 \msun, about 10\%
of the stellar mass.  In contrast, the circumbinary disk in UY Aur is resolved
only in CO; its continuum emission is compact and is therefore presumed to be
associated with only one of the binary components.  

CO observations
show both circumbinary disks clearly separated from circumstellar disks by
regions of very low surface density, indicative of disk clearing. 
Furthermore, the circumbinary disks
have inner radii consistent with theory under reasonable assumptions about the
binary orientation.
The near-infrared scattered-light adaptive optics images also suggest 
the presence of radial structures in these gaps that could
arise from small amounts of infalling material; however, in the case of GG Tau
these structures are not confirmed by optical HST images (J. Krist,
personal communication).  A third example of a directly detected circumbinary 
disk is found in an optical scattered light image of 
the main-sequence B5 star, BD+31$^\circ$643, which has a projected binary
star separation of $\sim$200 AU 
(Kalas and Jewitt 1997).
 
As noted in Section IV.A., many PMS spectroscopic binaries have strong millimeter
emission that, given their small semimajor axes, requires
the presence of massive circumbinary disks. 
High spatial resolution interferometric maps of UZ Tau E (Fig.\ 4)
resolve the millimeter emission, confirming that a massive disk surrounds the
binary. Mid-infrared emission further
reveals the presence of circumbinary material around yet more close binaries.
The measured masses of the circumbinary
disks of DQ Tau (0.02 \msun), UZ Tau E (0.06 \msun) and GW Ori (0.3 \msun) all
exceed the minimum mass of the solar nebula. All have
semimajor axes $\simlt 1$ AU, and therefore these systems could have
stable planetary orbits in their circumbinary disks at distances as small as
$\approx 3$ AU\null.

Such close binary
systems are also excellent probes of disk clearing. The innermost regions of
disks are, as yet, inaccessible to imaging at the distances of the nearest
star-forming regions. However, these hot inner disks
contribute essentially all of the near-infrared ($\lambda
\simeq 2.2$--5 \micron) excess emission in the spectral energy
distributions of young stellar objects.  Thus, the lack of a near-infrared
excess indicates a lack of hot disk material.  
Jensen and Mathieu (1997)
studied the spectral energy distributions of all known young spectroscopic
binaries with disks in order to search for disk
clearing.  Indeed, some of the binaries (V4046 Sgr, 162814$-$2427)
have {\it no\/} near-infrared excess emission but substantial mid- and
far-infrared excesses, the signature of cleared inner disks.  The
inferred sizes of these inner holes are consistent with the sizes
expected given the binary orbits.  
\footnote{$^\dagger$}{It is worth noting that care must be taken in
interpreting small
depressions in SEDs as evidence for cleared inner disks, particularly
when no known companion is present, as
these dips can also be caused by dust grain opacity and
vertical temperature structure effects in a circumstellar disk (Boss and Yorke
1993, 1996).}

Surprisingly, however, a number of
binaries (UZ Tau E, DQ Tau, AK Sco) show relatively smooth power-law
spectral energy distributions, as expected from a continuous accretion disk.
The near-infrared opacity of dust is very large,
so that relatively little material is required to produce the observed
near-infrared emission.  This leaves open the possibility that these inner disks
have been dynamically cleared, but not with 100\% efficiency. Such a situation
could arise if material were to leak steadily from circumbinary disks into a
cleared gap.
Recent near-infrared CO observations of DQ Tau also reveal the presence
of hot gas near the stars (Carr et al., in preparation).

\subsection{C.~~Circumstellar disks}

The existence of circumstellar disks has always been implicit in the
discovery of binaries among CTTS, presuming that CTTS diagnostics are indeed
indicative of accretion at stellar surfaces. The outstanding issue is
whether circumstellar disks surround both primary and secondary stars, and
how their relative accretion rates compare.

These questions can be answered best through observations capable of
resolving the binary systems.  As an example, recent observations of
HK Tau by speckle imaging (Koresko 1998) and HST/WFPC2 (Stapelfeldt et
al.\ 1998) provide direct evidence of a circumsecondary disk in a wide
PMS binary star. (The HST image is shown in Fig.\ 1d of McCaughrean et
al., this volume.) In both observations the secondary is observed as
two elongated reflection nebulosities separated by a dark lane, well
matched to scattered light models of an optically thick circumstellar
disk seen close to edge-on.  The disk has a radius of $\sim 100$ AU,
roughly one-third the projected separation of the binary. Statistical
arguments for the true orbital elements suggest that dynamical
truncation of the circumsecondary disk is a possibility (Stapelfeldt
et al.).  Likewise, mid-infrared observations of the somewhat more
evolved A0 star HR 4796A reveal a circumprimary disk, still present at
an age of $\sim 8$ Myr despite the presence of a companion star
located 500 AU away (Koerner et al.\ 1998; Jayawardhana et al.\ 1998).
If the companion has an eccentric orbit and is currently near
apastron, it could influence the disk outer radius, but the observed
confinement of the disk material in a narrow annulus roughly 60--80 AU
from the primary is puzzling, perhaps suggesting the presence of one
or more unseen companions (Schneider et al.\ 1999).  An additional
influence of a distant companion star can be to warp the disk (Terquem
and Bertout 1993, Larwood et al.\ 1996).  Telesco et al.\ (1999) note that
their images of HR 4796 hint at a warp in the disk. Finally,
the secondaries of both HR 4796 and HK Tau show indirect evidence
of a circumstellar disk, and thus these systems appear to 
support both circumprimary and circumsecondary disks.  

Circumstellar disks in binary stars have also been identified
in observations that separate the emission for the
primary and secondary stars but do not resolve the individual disks,
using the same indirect measurements used to assess the presence of a
disk in single stars such as infrared and ultraviolet excesses and
strong emission lines.
For the widest binaries, separated by $\simgt 50$ AU,
Brandner and Zinnecker (1997), Prato and Simon (1997), Prato (1998), and
Duch\^ene et al.\ (1999b) have
investigated the occurrence of circumstellar disks through
a spectroscopic study of the individual components' emission line 
(H$\alpha$ and Br $\gamma$) characteristics. Of the combined 49 binaries
whose combined light has been classified as CTTS, 43 include two CTTS stars,
suggesting that they harbor both circumprimary and
circumsecondary disks.

Observations with the HST/WFPC2 and ground-based speckle im\-aging
have permitted investigations of circumstellar disks among 31 binaries
with separations of 10 AU to 100 AU (Ghez et al.\ 1997b; White and Ghez 1999) 
using the individual components' ultraviolet and infrared
excesses as proxies for the presence of circumstellar disks. 
These studies have also shown that for the majority of pairs with accretion
both components show similar accretion signatures. Interestingly, 
among the remainder (29\%) of
the active pairs only the primary retains an accreting disk, suggesting
that primaries may have somewhat longer lived disks. Moreover,
the excesses of the primaries are generally larger than
or comparable to those of the secondaries, indicating that the primary stars are
experiencing larger accretion rates. If primary disks do indeed have longer
lifetimes and higher accretion rates than do secondary disks, then the primary
disks must be more massive or preferentially replenished. However, near-infrared
excesses do not provide meaningful measures of disks masses, and so the specific
masses of primary and secondary disks are unknown. Nonetheless, the presence
of large accretion rates would suggest that massive
circumstellar disks may survive in close binary systems.

High spatial resolution interferometric observations allow the 
millimeter emission to be localized even in binary stars 
with separations $\simlt 100$ AU\null.  Three clear cases
of circumstellar disks based on their compact millimeter emission 
are GG Tau (Guilloteau et al. 1999), UZ Tau W
(Jensen et al.\ 1996b; see Fig.\ 4, Section IV.A), and T Tau N (Akeson et al.
1998).  In the case of GG Tau, Guilloteau et al.\ report that the emission is
consistent with tidally truncated circumstellar 
disks of radius $R \sim 4$--20 AU and
total mass $ \geq 1.5\times 10^{-4}\ \msun$.

Circumstellar disks may even exist in the closest binary star 
systems.  As discussed in Section V, PMS spectroscopic binaries can
show infrared excesses, high-amplitude photometric variability, 
H$\alpha$ equivalent 
widths in excess of 100 \AA, heavy veiling, and large ultraviolet excesses.  
These observations are indicative of accretion and material 
very near the stars, which may be in the form of either 
accretion disks or accretion flows. 

Finally, an important but relatively unknown property of circumstellar disks
is their spatial orientation in the binary system.  Disk alignment is one
of the few observable properties that can distinguish among competing
models of binary formation.  It is also critically important for the
long-term stability of planetary systems.  
However, the challenge of resolving
circumstellar disks has limited our knowledge of disk alignment.  A
notable exception is the HK Tau system (Stapelfeldt et al.\ 1998,
Koresko 1998) where the resolved disk appears not to be coplanar with
the binary orbital plane.  Disk alignment also can be probed by polarimetry
even in systems where the disks themselves are unresolved.  Scattering
off a disk introduces a net polarization, and so the polarization
position angle (PA) of the combined light from disk and star indicates
the PA of the disk on the sky
(Koerner and Sargent 1995).  Monin et al.\ (1998)
compiled polarimetric measurements of 8\arcsec--37\arcsec\ binaries
from the literature and measured the polarization of some closer
systems.  They found one system (GI/GK Tau) to have mis-aligned
polarization vectors, while two other systems were consistent with
being parallel.  Jensen and Mathieu (in preparation) have made K-band
imaging polarimetric measurements of all $>1$\arcsec\ young binaries
with infrared excess in Taurus and Ophiuchus.  Preliminary results
show mis-aligned disks in HK Tau (with both components detected,
indicating an unresolved disk around the primary) and in DK Tau.
Currently this technique is limited to binaries wider than about
1\arcsec\ due to the lack of sensitive, stable polarimeters combined
with adaptive optics, so it cannot probe binaries with separations
much less than typical disk radii to see if the disks in
closer binaries are aligned.  A change in alignment properties with binary
separation may shed light on whether there are different formation
mechanisms for the closer and wider binaries. 

\mainsection{{V}.~~DISKS IN YOUNG BINARY STARS: ACCRETION\hfill \break\indent FLOWS}

The observation of accretion diagnostics in CTTS binaries of {\it all}
separations suggests that accretion continues at stellar surfaces, with the
accreting material presumably flowing from circumstellar disks.
However, the theoretical expectation has been that
the balance of viscous and resonant forces at the inner edge of a circumbinary
disk would prevent any flow of circumbinary material across the gap. The
consequence of continued accretion at the stellar surfaces would thus
ultimately be exhaustion of the circumstellar disks and the cessation of
accretion. Thus one might expect the accretion timescale for binaries to
differ from that of single stars, and indeed to differ between close and
wide binaries.
In fact there is little observational evidence for this; for example,
Simon and Prato (1995)
find no difference in the frequency of accretion diagnostics between single
and binary stars. Such
long-lived accretion from
circumstellar disks is an outstanding puzzle, one which may lead to a much more
dynamic view of disk evolution.

The widest binaries ($>$100 AU) whose circumstellar masses appear to be 
unaffected by their distant companions (Section IV.A) are likely to 
have accretion histories very similar to single stars.
Thus the existence of active accretion in wide binaries
is not a surprise, where here we operationally define ``wide'' as a separation
several times greater than typical disk radii. More surprising, perhaps, is
an apparent correlation in the presence of accretion onto primary and secondary
stars (see above discusion of Brandner and Zinnecker (1997), Prato and Simon (1997),
Prato (1998), Duch\^ene et al.\ (1999b), White and Ghez (1999)).
These studies suggest that the components
of binary systems typically retain their disks for similar lengths of time.
Prato and Simon suggested that a
common circumbinary envelope may replenish and maintain
circumstellar disks around both stars, although evidence for such envelopes
has not been found for many of the binaries in these studies.

In their HST study of four CTTS and two WTTS with separations of
10 AU to 50 AU, Ghez et al.\ (1997b) also
found that both components of three
close CTTS binaries (including UZ Tau W) show infrared excesses
suggestive of circumstellar disks, and that two stars have measurable
ultraviolet excesses indicative of active accretion. 
The ultraviolet excesses from the two stars 
suggest that high accretion rates continue
even in binary systems with separations much less than typical disk radii.

Perhaps most remarkably, the observational diagnostics for accretion
are present among even the very closest binary stars. The star UZ Tau E is
arguably one of the most classic of CTTS\null. It is a high-amplitude photometric
variable (and indeed noted as one of the most active T Tauri stars by
Herbig (1977)), has an emission
spectrum with H$\alpha$ equivalent widths in excess of 100 \AA, is often
heavily veiled, and has a large ultraviolet excess. Analyses have suggested
accretion rates as high as $2 \times 10^{-6}$  \msun/yr 
(Hartigan et al.\ 1995). 
In addition it is surrounded by a massive
disk which has been resolved at millimeter wavelengths, and it has a power-law spectral
energy distribution with excess emission at all infrared wavelengths.
Despite these paradigmatic diagnostics for a disk accreting onto a single PMS 
star, UZ Tau E is a spectroscopic binary with a period of 19 days 
(Mathieu et al.\ 1996). Furthermore, with a maximum periastron separation of 0.1 
AU, it is
clear that the suggested accretion rate cannot be fed solely by an 
unreplenished circumstellar disk, although there is ample material in the 
circumbinary disk.

UZ Tau E is not a unique case. The CTTS AK Sco and DQ Tau have similar orbital
periods and show diagnostics of accretion and material very near the stars.
Indeed, the infrared spectral energy distribution of DQ Tau is one of the best 
examples among CTTS of a power law (Mathieu et al.\ 1997), typically interpreted as a 
continuous disk. Like UZ Tau E, both of these binaries have massive circumbinary
disks. The issue is whether the circumbinary material can be tapped to supply
the material accreting onto the stellar surfaces. 
Photometric and spectroscopic monitoring of DQ Tau may have provided a 
clue to the tapping mechanism. Photometric monitoring has revealed periodic 
brightenings with a period of 15.8 days. This is precisely the same as the orbital
period, with the brightenings occurring at periastron passage. During these
brightenings the system becomes bluer, the veiling increases, and emission line
strengths increase (Mathieu et al.\ 1997; Basri et al.\ 1997). 
Together, all of these results point toward an
increased mass accretion rate at periastron passage.

Basri, Mathieu and collaborators have argued that these results
are consistent with the presence of accretion streams from the circumbinary
disk to at least one of the stellar surfaces. Recent
theoretical work of Artymowicz and Lubow (1996) and Lubow and Artymowicz (this
volume) has suggested
that such streams may develop if the circumbinary disks are sufficiently
viscous and warm. 
In this scenario, the increase in luminosity is due to the deposition of kinetic
energy from the infalling streams and the consequent heating of a region near 
the stars. For a binary with elements similar to DQ Tau, their
simulations argue that the streams will be pulsed, with maximum accretion
rates at periastron passage, as observed. Clearly the next
observational forefront in the study of disks in binary systems is angular
resolution on scales much smaller than companion separations, so that the
dynamic environment within the binary orbit can be imaged.

\mainsection{{V}{I}.~~L1551 IRS5: OPENING THE WINDOW ON PROTOBINARIES}
An exciting observational frontier lies in the direction of binaries in
formation. Several very wide pairs have been found among embedded sources.
However, perhaps the case of L1551 IRS5 best shows the prospects in the 
application of forefront technology to embedded sources.

L1551 IRS5 is arguably the canonical Class I system,  i.e.\ a protostar still
undergoing infall from an envelope.
It has long been known to be double at centimeter wavelengths, but
the interpretation as a binary system has not been secure. Recent millimeter
observations demonstrate that the system is a protobinary with
a projected separation of 45 AU (Looney et al.\ 1997, Rodriguez et al.\ 1998).
The 7-mm VLA observations with 7 AU resolution are particularly impressive
(Rodriguez et al.\ 1998). As shown in Fig.\ 6, not only is the binary clearly
evident but the circumstellar emissions are marginally resolved as well. That
these are in fact circumstellar disks is indicated by their flux levels being
well above the extrapolation of the centimeter spectral energy distribution
and by their orientation perpendicular to the centimeter outflow emission.

Combining the wealth of observations of L1551 IRS5, Looney et al.\ (1997)
describe the system as having three main components: a large-scale envelope,
a disk or extended structure with a size scale of order 150 AU, and the inner
binary system with two circumstellar disks. They suggest an envelope mass
and inner and outer radii of roughly 0.44 \msun, 30 AU and 1300 AU, 
respectively. The nature
of the circumbinary structure remains uncertain. First resolved by Lay et al.\
(1994), the emission is reasonably fit with a Gaussian model with a very
rough mass of 0.04 \msun. The VLA observations indicate circumstellar disk radii
of 10 AU, with disk masses of 0.06 \msun\ and 0.03 \msun. These disk masses are
uncertain due to contamination by free-free emission, but Rodriguez et al.\
argue that this overestimate is no more than a factor of 4. As such, the
disk masses remain comparable with the minimum mass required to form a
planetary system like our own.

It is notable that such large circumstellar masses are found in a system with
projected separation of 45 AU, given that at later evolutionary stages such
binaries typically do not have detectable millimeter flux (Section IV.A).
Indeed, given radii of only 10 AU the inferred disk masses are quite large.
Perhaps these substantial circumstellar disks are maintained
by rapid accretion from the envelope, possibly via an accretion stream
from a circumbinary disk.

\mainsection{{V}{I}{I}.~~CLOSING THOUGHTS ON PLANET FORMATION}
A wide variety of investigations in the last several years have made
it clear that disks are common in binary systems of all separations.
There is no significant difference in the frequency of near-infrared
excess emission in binaries and single stars (Moneti and Zinnecker 1991,
Simon and Prato 1995,
White and Ghez 1999),
nor in the frequency of detected {\it IRAS\/} 12, 25, or 60 \micron\
emission (Jensen et al.\ 1996a).  This leaves open the very real
possibility that while disks in binary systems may be constrained in
{\it size\/} by the presence of a companion, the remaining circumstellar
disk material may be similar in temperature and surface density to
that in disks around single stars.  The clear implication is
that in a significant fraction of binary systems planet formation may
be able to proceed relatively undisturbed. Here we look more closely at
exactly which binaries may be likely or unlikely to form planetary
systems based on our current knowledge.

In our own solar system, most of the planet formation has occurred in
the range of 0.1--30 AU\null.  If this holds true in other systems,
what fraction of binaries could form similar planetary systems?  A
typical circumstellar disk radius is predicted by theory to be no more than
0.3 times the
binary semimajor axis, allowing 30 AU disks in any binary wider than
about 100 AU\null.  Adopting the period distribution for main-sequence
G stars (Duquennoy and Mayor 1991), this includes roughly 37\% of all
binary systems.  Narrowing the disk radius to 5 AU (i.e., including only
the terrestrial planets plus the asteroid belt) reduces the minimum
binary separation to 17 AU and includes 57\% of all binaries.  These
numbers assume that planet formation is a relatively local event,
proceeding independently of conditions elsewhere in the system, which may
not be the case. Nonetheless, it
is clear that the possibility of planet formation in even relatively
close binary systems is by no means ruled out by current data.  These
numbers also do not include the roughly 10\% of very small separation
binaries ($a \simlt 0.3$ AU) with circumbinary disks that begin at or
inside 1 AU; such systems could form circumbinary planetary systems.

In wider (hundreds of AU) systems, the question has already been
answered.  Four of the eight extrasolar planetary systems known to
date are in members of wide binary systems, with binary separations
ranging from $\sim 200$--1000 AU\null (Marcy et al./, this volume).

Finally, we note that more subtle effects may be important for the
long-term prospects of planets in binary systems.  Little is currently
known about the relative alignment of circumstellar disks and the
binary orbital plane.  Holman and Wiegert (1996; see also Wiegert and
Holman 1997, Innanen et al.\ 1997) 
found that a planetary system around one member of a wide
binary is stable for longer times if the planetary orbits and binary
orbit are co-planar; in their calculations, planetary systems
inclined to the binary orbit become unbound in $10^7$--$10^8$ yr,
while a coplanar planetary system survives for the full $10^9$ yr
integration.  
Investigations of disk alignment in binary systems
will be important in determining the long-term stability of planets in
binaries and therefore their prospects for evolving and supporting
life.
\vskip 0.2in

We would like to gratefully acknowledge assistance with figures provided
by A. Dutrey, L. Rodriguez, and R. White. RDM was supported in part by
NSF grant AST-941715. Funding for AMG's contribution was provided by
NAG5-6975. 
ELNJ and RDM were supported by the NSF's Life in Extreme Environments program
through grant AST 97-14246.
MS was supported by NSF grant AST 94-17191.

\vfill\eject
\input psfig.sty
\hsize=5.25truein
\hoffset=0.5truein
\def\plotone#1{\psfig{file=#1,width=5.25truein}}
\count20=1
\def\caption#1{\noindent Figure \number\count20 .  #1
\advance\count20 by 1
\vfil\eject}
\plotone{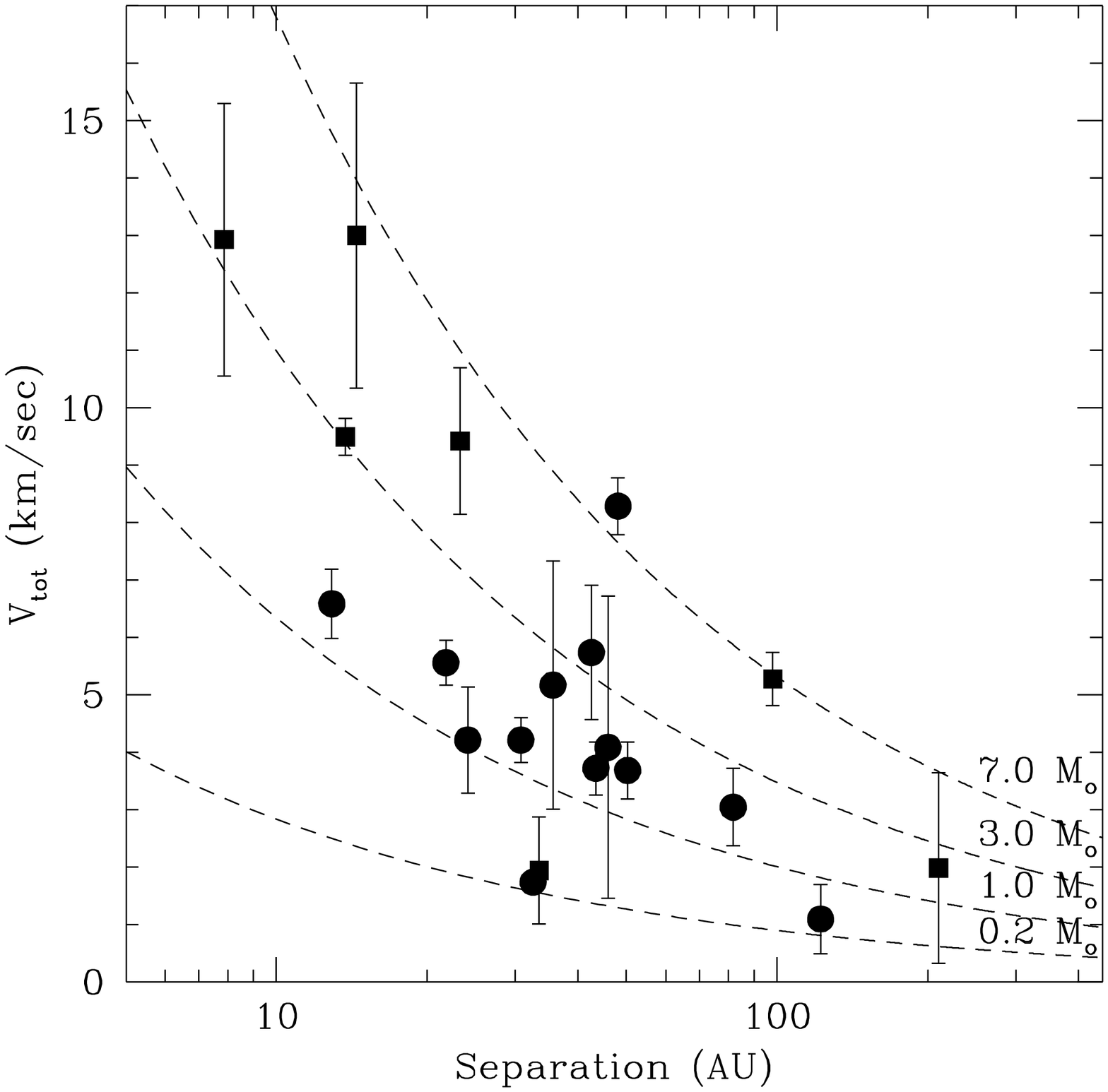}
  \caption{
The relative velocity of the binary stars' components as a function
of their mean separation.  The observed velocities are consistent with orbital
motion since they (1) decrease with separation and (2) are generally greater for
systems with higher mass primary stars ($M_1 >  1  M_{\odot}$ and $M_1 \le  1  M_{\odot}$ are plotted as squares and circles respectively).
The measurements are compared to that expected from a set of
randomly oriented binary stars
with total masses of 0.2, 1, 3, and 7 $M_{\odot}$.  
Although 
any individual total mass estimate is unreliable due to projection effects,
the sample has an average
total dynamical mass of $\sim$1.7 $M_{\odot}$.
(Taken from Ghez et al.\ 1995.)}

\plotone{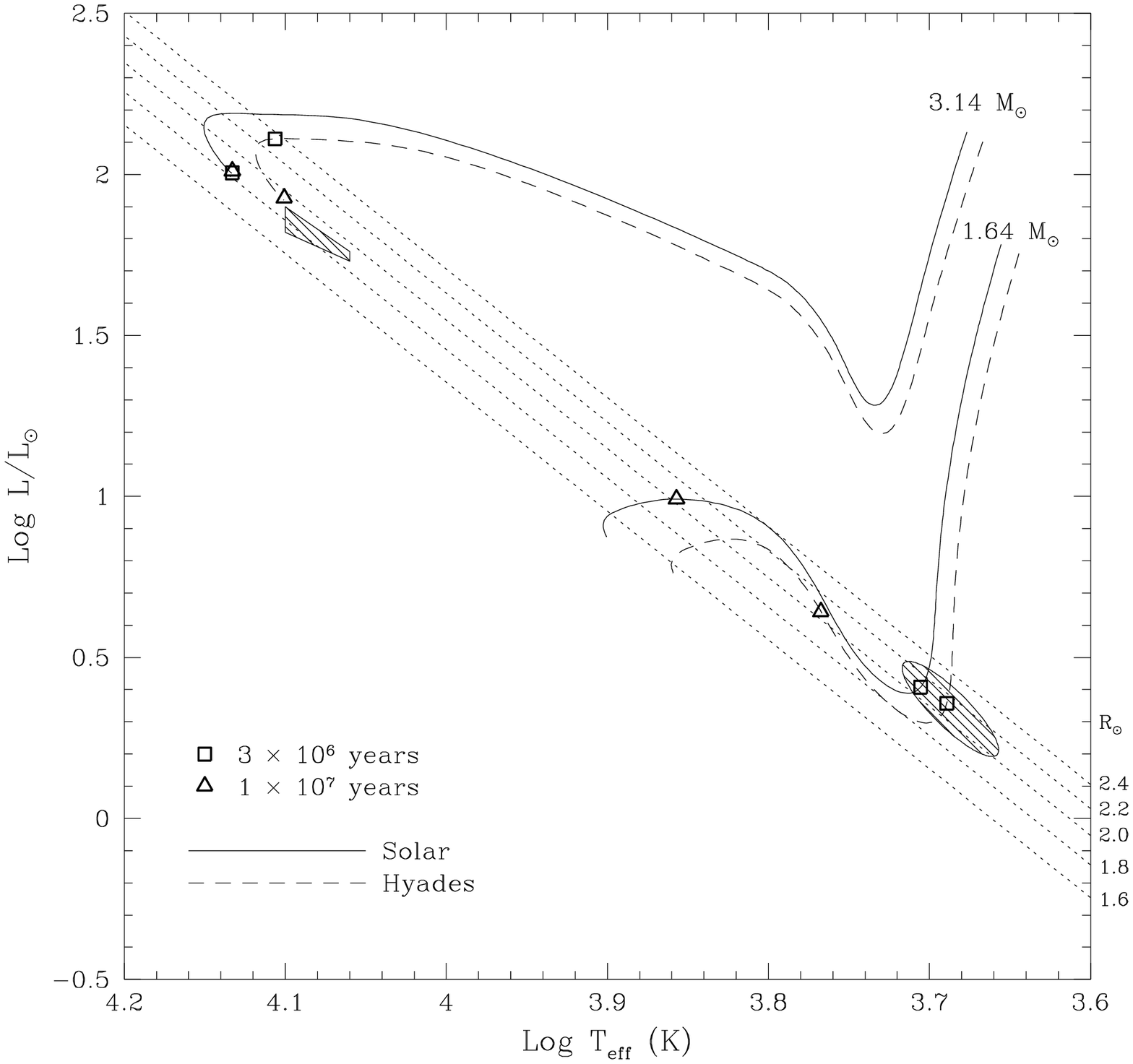}
\caption{
Location of TY~CrA primary and secondary stars in the theoretical H-R diagram.
The hatched regions designate high-confidence domains for the primary and
secondary based on light-curve analyses. Dotted lines are drawn at constant
radii. Solid and dashed lines correspond to pre-main-sequence tracks of
Swenson et al.\ (1994), calculated for the masses of the TY~CrA components at
solar (solid curve) and Hyades (dashed curve) compositions. Open boxes and
triangles mark isochrone points at ages of $3 \times 10^6$ yr and
$10 \times 10^7$ yr, respectively.
(Taken from Casey et al.\ 1998.)}

\plotone{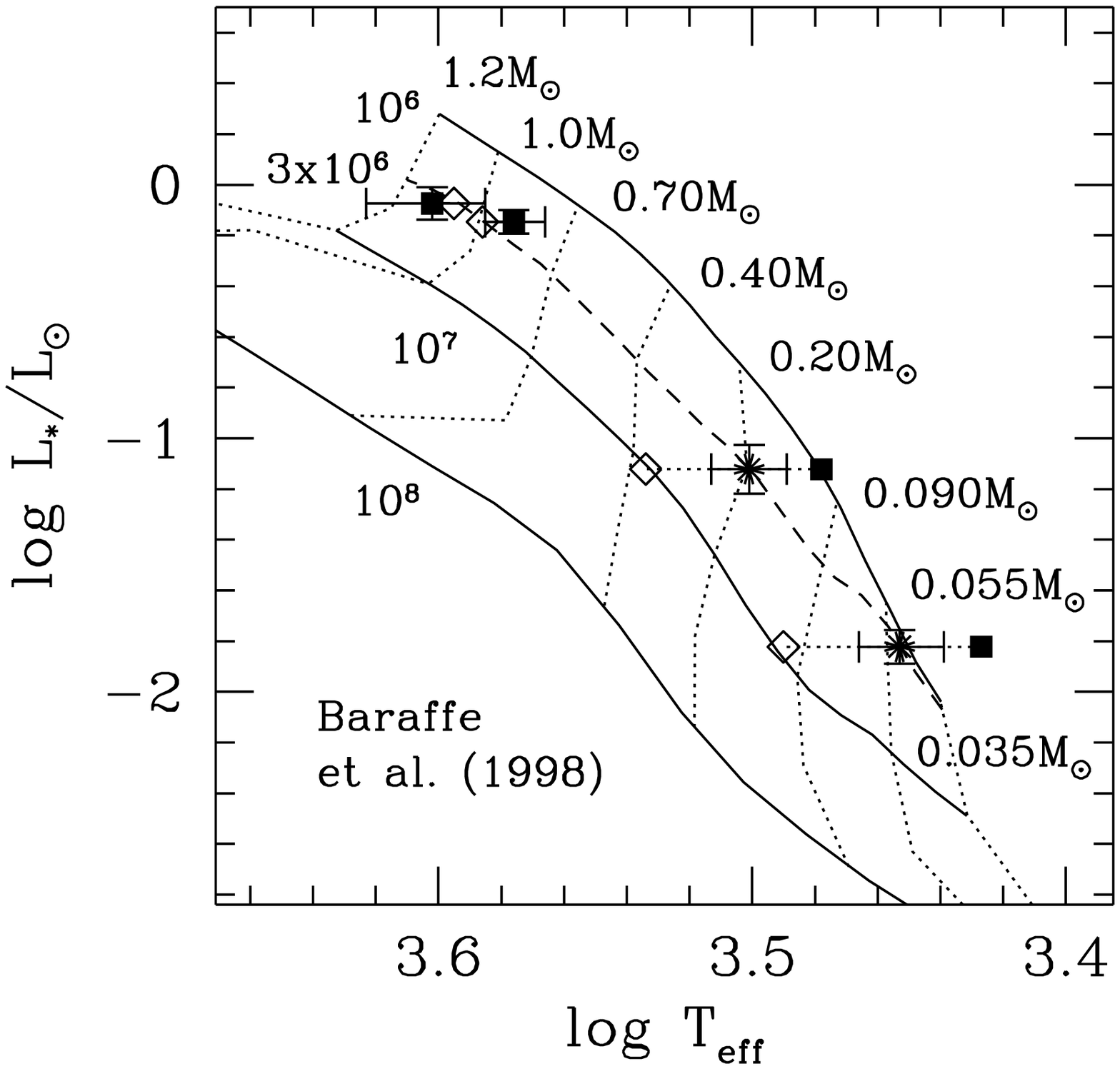}
\caption{
The stars of the GG Tau quadruple system compared with the theoretical 
evolutionary tracks of Baraffe et al.\ (1998; $\alpha=1.9$).
The range of plausible temperatures
for each component are determined using a dwarf temperature scale
(solid squares; Leggett et al.\ 1996) and a giant temperature scale
(open diamonds; Perrin et al.\ 1998).  Since the dwarf and giant
temperature scales are nearly identical for the two hottest components,
these two stars define an isochrone (dashed line) that can be used
to test evolutionary models and the T Tauri temperature scale
at lower masses.  Of the models tested, the Baraffe et al.\ 
models yield the most consistent ages using a temperature scale
intermediate between that of dwarfs and giants.  These tracks and the
implied coeval temperature scale (asterisks) yield a substellar mass of
0.044 $\pm$ 0.04  M$_\odot$ for the lowest mass component of GG Tau. 
(Taken from White et al.\ 1999)}

\plotone{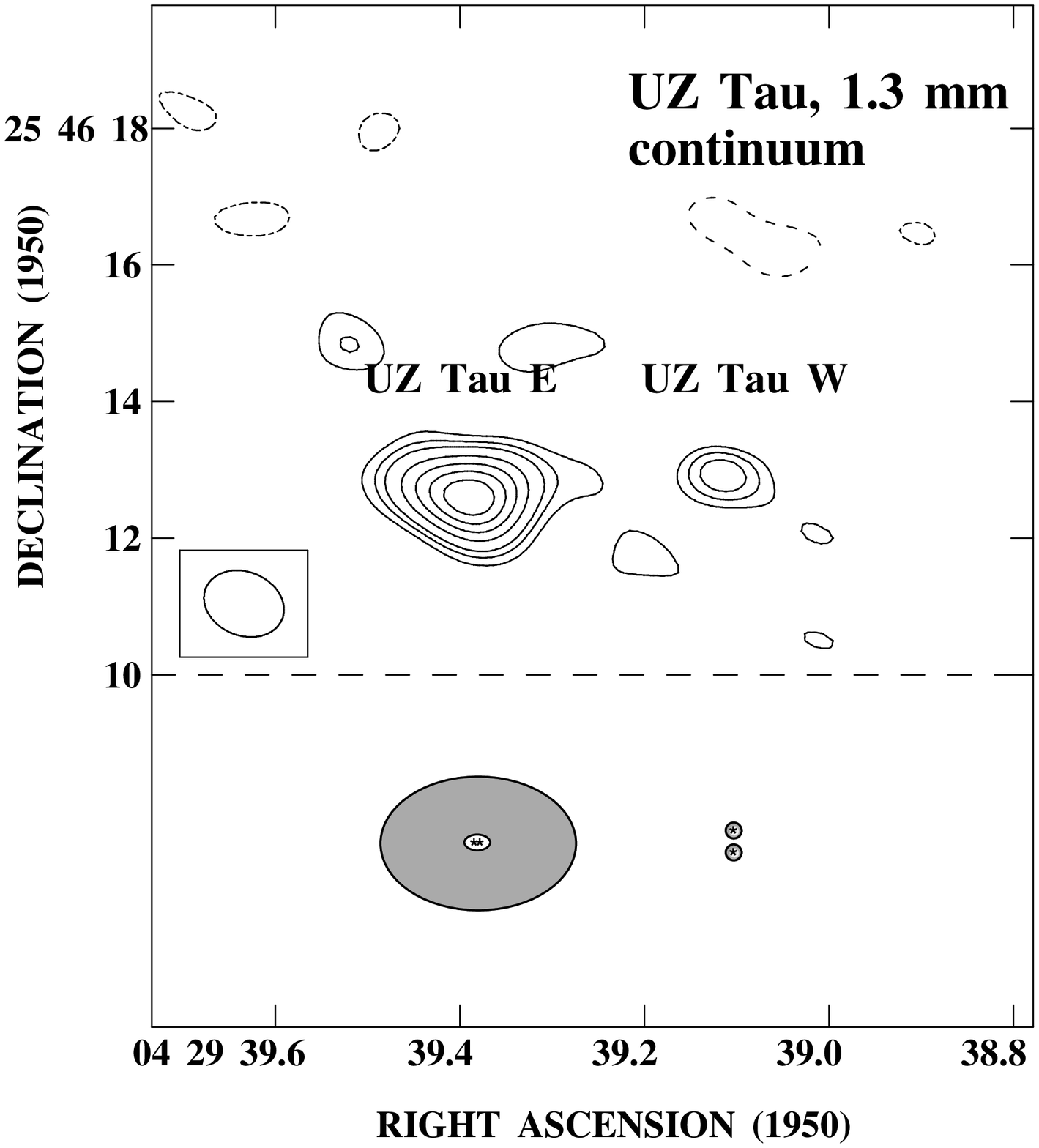}
\vskip 3mm
\caption{
    Owens Valley Radio Observatory $\lambda = 1.3$ mm continuum map (top) of
    UZ Tau, a young quadruple system, along with a schematic model for the
system (bottom). UZ Tau W is a 50 AU binary,
    separated by 500 AU from the sub-AU binary UZ Tau E\null.
    UZ Tau E has a massive circumbinary disk, seemingly undisturbed by either
    the central close binary or by UZ Tau W located several disk
    radii away in projection. In contrast UZ Tau W, a binary with a
    separation similar to typical disk radii, shows substantially less disk 
    emission, though it retains {\it some\/} mass in one or two unresolved
    circumstellar disks. (Adapted from Jensen et al.\ 1996b.)}

~\vskip 1cm
\psfig{file=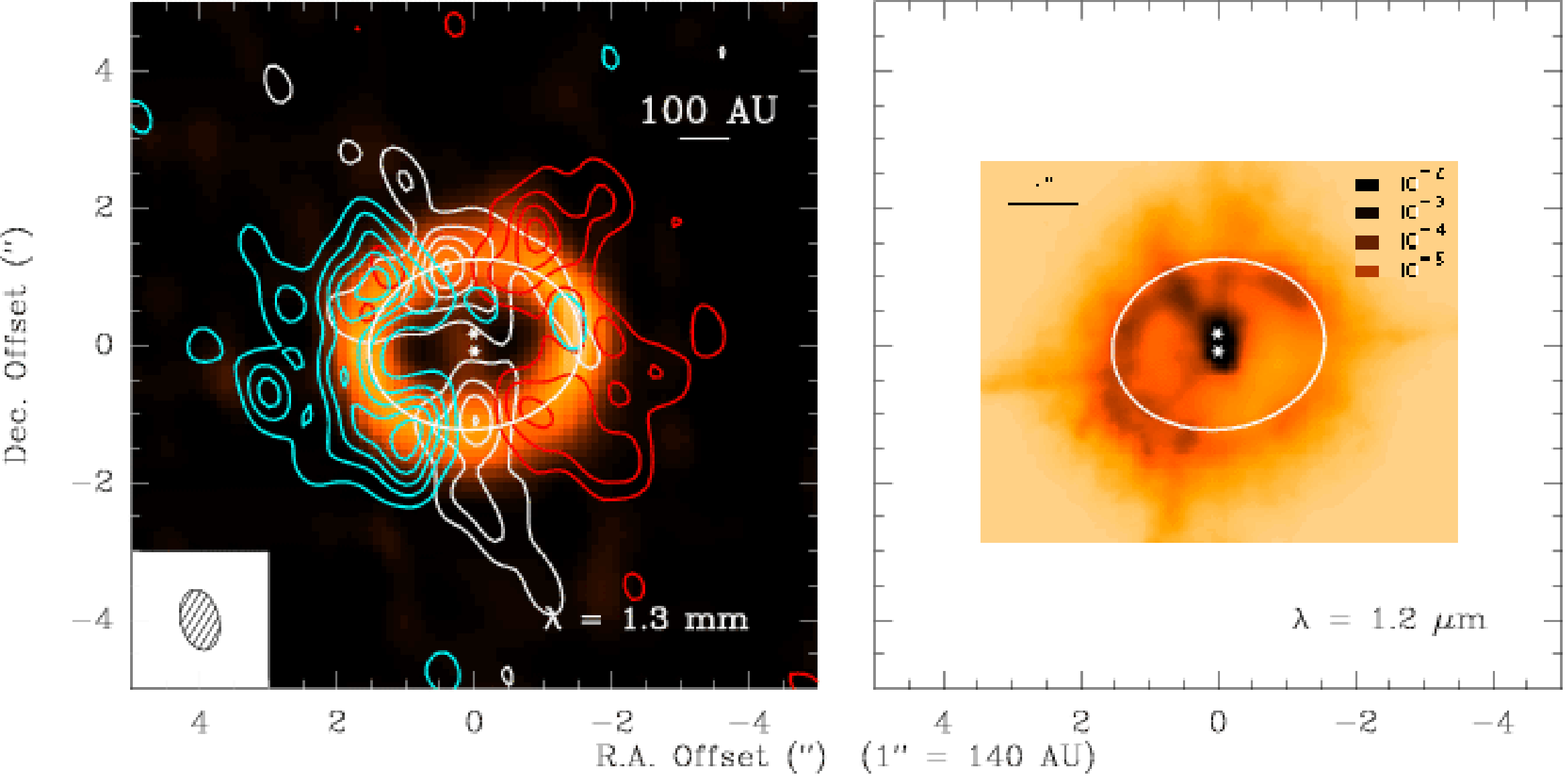,width=5.25truein,angle=270}
\vskip 1cm
\caption{
Left---The stellar components of the GG Tau binary are shown as stars.
The circumbinary ring is evident in the light shaded region,
displaying 1.3 mm continuum emission. 
The contours display three velocity channels of
$^{13}$CO 2$\rightarrow$1 emission: 5.55 km/sec, 6.30 km/sec and 7.05 km/sec, increasing
from left to right in the figure. The spatial gradient in the line emission
is consistent with Keplerian motion about a binary of 1.3 (D/140 pc) \msun. The
synthesized beam is $0.88''\times 0.56''$. Right---The J-band adaptive optics
image. The white ellipse is the same in both figures and represents the ring
average radius. Evidently the light in the near infrared coincides
with the circumbinary ring seen in millimeter light, and is interpreted as
scattered light off the inner edge of the ring. (Taken from Guilloteau
et al. 1999; the J-band image is from Roddier et al.\ 1996.)}

\psfig{file=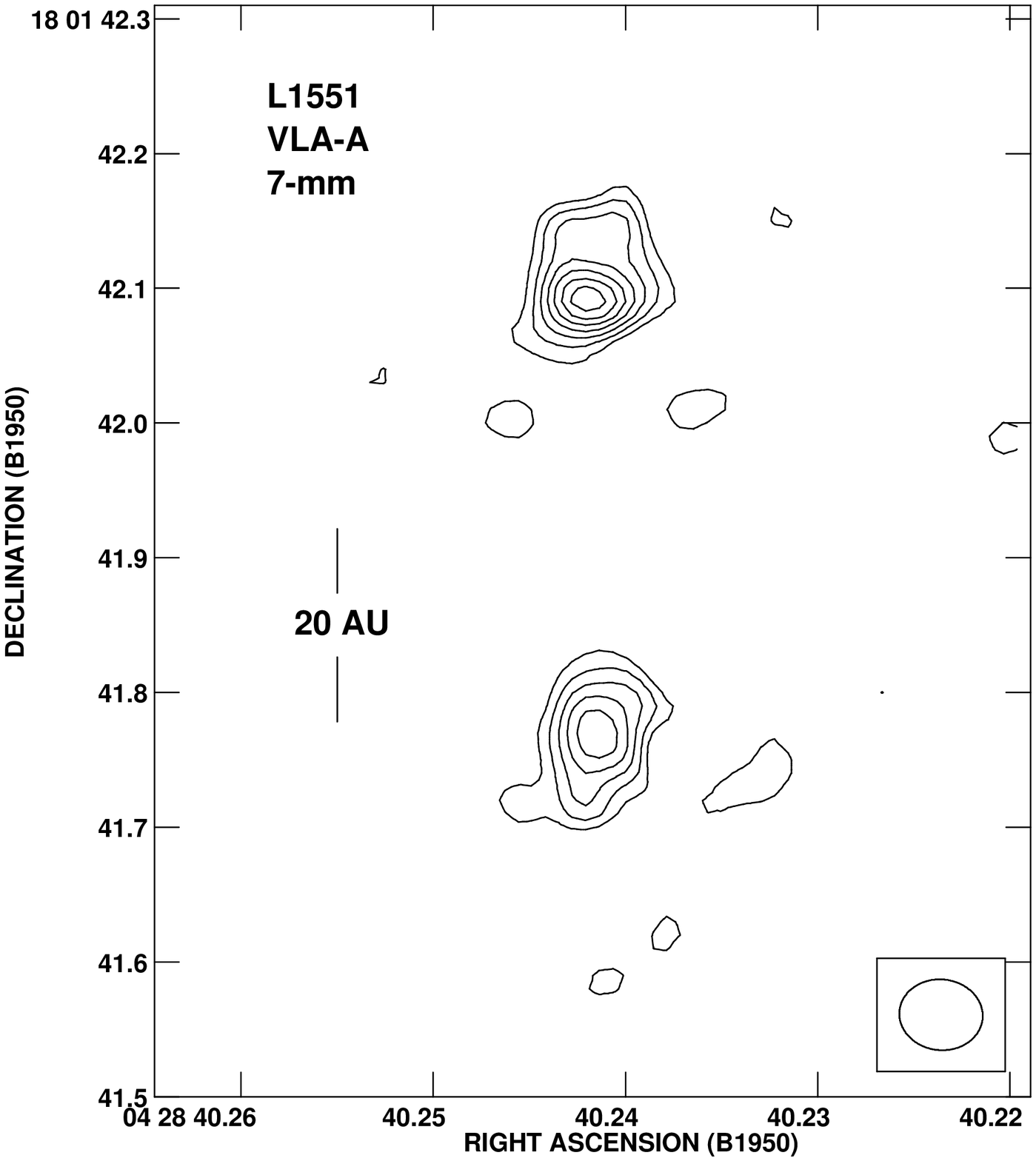,width=5.05truein}
\caption{
VLA image of the L1551 IRS5 region at 7 mm. The two compact sources are
interpreted as protoplanetary disks in a gravitationally bound
protobinary system. The masses of the individual disks are of
order 0.05 \msun, adequate to form planetary systems like our own.
However the spatially resolved projected semimajor axes are only 10 AU,
perhaps due to dynamical truncation by the stars. The half-power contour
of the VLA beam was $\approx 0.05 \arcsec$, opening a new forefront in 
high-resolution observations of young binary systems.
(Taken from Rodriguez et al.\ 1998.)}

\vfill\eject
\hsize=4.25truein
\hoffset=1truein
\vskip 0.5in
\centerline{\bf REFERENCES}
\vskip .25in

\def\araa{{\refit Ann.\ Rev.\ Astron.\ Astrophys.\/}}
\def\baas{{\refit Bull.\ American Astron.\ Society\/}}
\def\aj{{\refit Astron.\ J.\/}}
\def\apj{{\refit Astrophys.\ J.\/}}
\def\apjs{{\refit Astrophys.\ J.\ Suppl.\/}}
\def\apjl{{\refit Astrophys.\ J.\ Lett.\/}}
\def\aap{{\refit Astron.\ Astrophys.\/}}
\def\mnras{{\refit Mon.\ Not.\ Roy.\ Astron.\ Soc.\/}}
\def\ppiii{{\refit Protostars \& Planets {I}{I}{I}\/}, eds.\ E. H. Levy and
J. I. Lunine (Tucson: Univ.\ of Arizona Press)}

\ref{Akeson, R. L., Koerner, D. W., Jensen, E. L. N. 1998. 
A Circumstellar Dust Disk around T Tau N: Subarcsecond Imaging
at $\lambda$ = 3 millimeters.  \apj, 505:358-362} 
 
\ref{Artymowicz, P. \& Lubow, S. H. 1996.  Mass Flow through Gaps in 
Circumbinary Disks.  \apjl, 467:L77--L80.} 

\ref{Artymowicz, P. \& Lubow, S. H. 1994.  Dynamics of binary-disk 
interaction. 1: Resonances and disk gap sizes.  \apj, 421:651--667.}

\ref{Baraffe, I., Chabrier, G., Allard, F., \& Hauschildt, P. H. 1998.  
Evolutionary models for solar metallicity low-mass stars:
mass-magni-\hfil\break  tude 
relationships and color-magnitude diagrams.  \aap, 337:403--412.} 

\ref{Basri, G., Johns-Krull, C. M., \& Mathieu, R. D. 1997.  The Classical 
T Tauri Spectroscopic Binary DQ Tau. II. Emission Line Variations with 
Orbital Phase.  \aj, 114:781--792.} 

\ref{Beckwith, S. V. W. \& Sargent, A. I. 1993.  The occurrence and 
properties of disks around young stars. In \ppiii, pp.\ 521--541.}

\ref{Beckwith, S. V. W., Sargent, A. I., Chini, R. S., \& G\"usten, R. 1990. 
 A survey for circumstellar disks around young stellar objects.  \aj, 
99:924--945.}

\ref{Boss, A. P. \& Yorke, H. W. 1993.  An alternative to unseen companions 
to T Tauri stars.  \apjl, 411:L99--L102.}

\ref{Boss, A. P. \& Yorke, H. W. 1996.  Protoplanetary Disks, Mid-Infrared 
Dips, and Disk Gaps.  \apj, 469:366--372.} 

\ref{Bouvier, J., Rigaut, F., \& Nadeau, D. 1997.  Pleiades low-mass 
binaries: do companions affect the evolution of protoplanetary disks?  
\aap, 323:139--150.}

\ref{Brandner, W., Alcala, J. M., Kunkel, M., Moneti, A., \& Zinnecker, H. 1996.  
Multiplicity among T Tauri stars in OB and T associations.  Implications
for binary star formation.  \aap, 307:121--136.}

\ref{Brandner, W. \& K\"ohler, R. 1998.  Star Formation Environments and the 
Distribution of Binary Separations.  \apjl, 499: L79--L82.}

\ref{Brandner, W. \& Zinnecker, H. 1997.  Physical properties of 90 AU to 
250 AU pre--main-sequence binaries.  \aap, 321:220--228.} 

\ref{Brice\~no, C., Hartmann, L. W., Stauffer, J. R., Gagn\'e, M., \& Stern, R. 
A. 1997.  X-Rays Surveys and the Post-T Tauri Problem.  \aj, 113:740--752.} 

\ref{Burrows, A., Hubbard, W. B., Saumon, D., \& Lunine, J. I. 1993.  An 
expanded set of brown dwarf and very low mass star models.  \apj, 
406:158--171.} 

\ref{Casey, B. W., Mathieu, R. D., Vaz, L. P. R., Andersen, J., \& 
Suntzeff, N. B. 1998.  The Pre--Main-Sequence Eclipsing Binary TY Coronae 
Australis: Precise Stellar Dimensions and Tests of Evolutionary Models.  
\aj, 115:1617--1633.}

\ref{Claret, A. 1995.  Stellar models for a wide range of initial chemical 
compositions until helium burning. I. From X=0.60 to X=0.80 for Z=0.02.  
{\refit Astron.\  Astrophys.\ Suppl.\/}, 109:441--446.} 

\ref{Claret, A., Gimenez, A., \& Martin, E.L. 1995.  A test case of stellar 
evolution: the eclipsing binary EK Cephei. A system with accurate 
dimensions, apsidal motion rate and lithium depletion level.  \aap, 
302:741--744.}

\ref{Close, L. M., Dutrey, A., Roddier, F., Guilloteau, S., Roddier, C., 
Northcott, M., Menard, F., Duvert, G., Graves, J. E., \& Potter, D. 1998.  
Adaptive Optics Imaging of the Circumbinary Disk around the T Tauri Binary 
UY Aurigae: Estimates of the Binary Mass and Circumbinary Dust Grain Size 
Distribution.  \apj, 499:883--888.} 

\ref{Corporon, P., Lagrange, A.M., \& Beust, H. 1996.  Further 
characteristics of the young triple system TY Coronae Austrinae.  \aap, 
310:228--234.} 

\ref{D'Antona, F. \& Mazzitelli, I. 1994.  New pre--main-sequence tracks for 
$M \le 2.5$ \hbox{$M_{\odot}$}\ as tests of opacities and convection 
model.  \apjs, 90:467--500.}

\ref{Duchene, G., Bouvier, J. \& Simon, T. 1999a. Low-Mass binaries in the young
cluster IC348: Implications for binary formation and evolution. \aap, in press.}

\ref{Duch\^ene G., Monin J.-L., Bouvier J., M\'enard F. 1999b. Differential 
accretion in Taurus PMS binaries. \aap, submitted.}

\ref{Duquennoy, A. \& Mayor, M. 1991.  Multiplicity among solar-type stars 
in the solar neighbourhood. II - Distribution of the orbital elements in an 
unbiased sample.  \aap, 248:485--524.} 

\ref{Durisen, R.H. \& Sterzik, M.F. 1994.  Do star forming regions have 
different binary fractions?  \aap, 286:84--90.} 

\ref{Dutrey, A., Guilloteau, S., Duvert, G., Prato, L., Simon, M., 
Schuster, K., \& Menard, F. 1996.  Dust and gas distribution around T Tauri 
stars in Taurus-Auriga. I. Interferometric 2.7mm continuum and
$^{13}$CO $J=1\rightarrow 0 $
observations.  \aap, 309:493--504.} 

\ref{Dutrey, A., Guilloteau, S., Prato, L., Simon, M., Duvert, G., Schuster, K.,
\& Menard, F. 1998. CO study of the GM Aurigae Keplerian disk. \aap, 338, 
L63--66.}

\ref{Dutrey, A., Guilloteau, S., \& Simon, M. 1994.  Images of the GG Tauri 
rotating ring.  \aap, 286:149--159.} 

\ref{Duvert, G., Dutrey, A., Guilloteau, S., Menard, F., Schuster, K., 
Prato, L., \& Simon, M. 1998.  Disks in the UY Aurigae binary.  \aap, 
332:867--874.} 

\ref{Figueiredo, J. 1997.  The pre--main-sequence spectroscopic binary
\break NTTS 162814$-$2427: models versus observations.  \aap, 318:783--790.}

\ref{Ghez, A. M., McCarthy, D. W., Patience, J. L., \& Beck, T. L. 1997a.
The Multiplicity of Pre-Main-Sequence Stars in Southern Star-forming
Regions.  \apj, 481:378--385.}

\ref{Ghez, A. M., Neugebauer, G., \& Matthews, K. 1993.  The multiplicity 
of T Tauri stars in the star forming regions Taurus-Auriga and 
Ophiuchus-Scorpius: A 2.2 micron speckle imaging survey.  \aj, 
106:2005--2023.} 

\ref{Ghez, A. M., White, R. J., \& Simon, M. 1997b.  High Spatial Resolution 
Imaging of Pre--Main-Sequence Binary Stars: Resolving the Relationship 
between Disks and Close Companions.  \apj, 490:353--367.} 

\ref{Ghez, A. M., Weinberger, A.J., Neugebauer, G., Matthews, K., \& McCarthy,
D. W., Jr. 1995. Speckle Imaging Measurements of the Relative Tangential
Velocities of the Components of T Tauri Binary Stars. \aj, 111:753--765.}

\ref{Guilloteau, S. \& Dutrey, A. 1998. Physical parameters of the Keplerian 
protoplanetary disk of DM Tauri. \aap, 339:467--476.}

\ref{Guilloteau, S., Dutrey, A., \& Simon, M. 1999. GG Tau: the Ring World. 
\aap, in press.}

\ref{Hall, S. M. 1997.  Circumstellar disc density profiles: a dynamic 
approach.  \mnras, 287:148--154.} 

\ref{Hartigan, P., Edwards, S., \& Ghandour, L. 1995.  Disk Accretion and 
Mass Loss from Young Stars.  \apj, 452:736--768.} 

\ref{Hartigan, P., Strom, K. M., \& Strom, S. E. 1994.  Are wide 
pre--main-sequence binaries coeval?  \apj, 427:961--977.}

\ref{Herbig, G. H. 1977.  Eruptive phenomena in early stellar evolution.  
\apj, 217:693--715.} 

\ref{Herbig, G. H., and Bell, K. R. 1988.  Third Catalog of
Emission-Line Stars of the Orion Population. {\refit Lick Obs.\
Bulletin}, 1111.

\ref{Holman, M. \& Wiegert, P. 1996.  Long-term Stability of Planets in 
Binary Systems.  AAS/Division of Planetary Sciences Meeting,
28:12.12.}

\ref{Innanen, K. A., Zheng, J. Q., Mikkola, S., \& Valtonen, M. J. 1997.  
The Kozai Mechanism and the Stability of Planetary Orbits in Binary Star 
Systems.  \aj, 113:1915--1919.}

\ref{Jayawardhana, R., Fisher, S., Hartmann, L., Telesco, C., Pina, R., \& 
Fazio, G. 1998.  A Dust Disk Surrounding the Young A Star HR 4796A.  \apjl, 
503:L79--L82.}

\ref{Jensen, E. L. N., Koerner, D. W., \& Mathieu, R. D. 1996b.  
High-Resolution Imaging of Circumstellar Gas and Dust in UZ Tauri: 
Comparing Binary and Single-Star Disk Properties.  \aj, 111:2431--2438.}

\ref{Jensen, E. L. N. \& Mathieu, R. D. 1997.  Evidence for Cleared Regions 
in the Disks Around Pre--Main-Sequence Spectroscopic Binaries.  \aj, 
114:301--316.} 

\ref{Jensen, E. L. N., Mathieu, R. D., \& Fuller, G. A. 1994.  A connection 
between submillimeter continuum flux and separation in young binaries.  
\apjl, 429:L29--L32.} 

\ref{Jensen, E. L. N., Mathieu, R. D., \& Fuller, G. A. 1996a.  The 
Connection between Submillimeter Continuum Flux and Binary Separation in 
Young Binaries: Evidence of Interaction between Stars and Disks.  \apj, 
458:312--326.} 

\ref{Kalas, P., \& Jewitt, D. 1997.  A candidate dust disk surrounding the
binary stellar system BD+31$^\circ$643. {\refit Nature\/}, 386:52-54.}

\ref{Koerner, D. W., Ressler, M. E., Werner, M. W., \& Backman, D. E. 1998. 
 Mid-Infrared Imaging of a Circumstellar Disk around HR 4796: Mapping the 
Debris of Planetary Formation.  \apjl, 503:L83--L86.} 

\ref{Koerner, D. W. \& Sargent, A. I. 1995.  Imaging the Small-Scale 
Circumstellar Gas Around T Tauri Stars.  \aj, 109:2138--2145.}

\ref{K\"ohler, R. \& Leinert, C. 1998.  Multiplicity of T Tauri stars in 
Taurus after {\it ROSAT\/}.  \aap, 331:977--988.} 

\ref{K\"ohler, R., Leinert, Ch., Zinnecker, H. 1998
Multiplicity of T Tauri Stars in different Star-Forming Regions.
Astronomische Gessell\-schaft Meeting Abstracts, 14:17.}

\ref{Koresko, C. D. 1998.  A Circumstellar Disk in a Pre-main-sequence Binary
Star. \apj 507:145-148.}

\ref{Kroupa, P. 1995.  Star cluster evolution, dynamical age estimation and 
the kinematical signature of star formation.  \mnras, 277:1522--1540.} 

\ref{Lada, E. A., Evans, N. J., Depoy, D. L., \& Gatley, I 1991. 
A 2.2 micron survey in the L1630 molecular cloud. \apj, 371:171-182.}

\ref{Larwood, J. D., Nelson, R. P., Papaloizou, J. C. B., \& Terquem, C. 
1996.  The tidally induced warping, precession and truncation of accretion 
discs in binary systems: three-dimensional simulations.  \mnras, 
282:597--613.} 

\ref{Lay, O. P., Carlstrom, J. E., Hills, R. E., \& Phillips, T. G. 1994.  
Protostellar accretion disks resolved with the JCMT-CSO interferometer.  
\apjl, 434:L75--L78.} 

\ref{Lee, C.-W. 1992.  Double-lined pre--main-sequence binaries: A
test of pre--main-sequence evolutionary theory.  Ph.D. Thesis, Univ.\
of Wisconsin-Madison.}

\ref{Leggett, S. K., Allard, F., Berriman, G., Dahn, C. C., Hauschildt, P. H.
1996.  Infrared Spectra of Low-Mass Stars: Toward a Temperature Scale for
Red Dwarfs. \apjs, 104:117-143.}

\ref{Leinert, C., Zinnecker, H., Weitzel, N., Christou, J., Ridgway, S. 
T., Jameson, R., Haas, M., \& Lenzen, R. 1993.  A systematic search for 
young binaries in Taurus.  \aap, 278:129--149.}

\ref{Looney, L. W., Mundy, L. G., \& Welch, W. J. 1997.  High-Resolution 
$\lambda = 2.7$ Millimeter Observations of L1551 IRS 5.  \apjl, 484:L157--L160.} 

\ref{Malkov, O., Piskunov, A., \& Zinnecker, H. 1998. On the luminosity ratio 
of pre-main sequence binaries. \aap, 338:452-454.}

\ref{Martin, E.L. \& Rebolo, R. 1993.  EK Cephei B: a test object for 
pre-ZAMS models of solar-type stars.  \aap, 274:274--278.} 

\ref{Mason, B.D., Henry, T.J., Hartkopf, W.I., Ten Brummelaar, T., \&
Soderblom, D.R. 1998. A Multiplicity Survey of Chromospherically Active and 
Inactive Stars. \aj, 116:2975-2983.}

\ref{Mathieu, R. D. 1994.  Pre--Main-Sequence Binary Stars.  \araa, 
32:465--530.} 

\ref{Mathieu, R. D. 1996. Binary Frequencies Among Pre--Main-Sequence
Stars.  In {\refit The Origins, Evolutions and Destinies of Binary
Stars in Clusters\/}, ASP Conference Series, 90, 231.}

\ref{Mathieu, R. D., Adams, F. C., Fuller, G. A., Jensen, E. L. N., 
Koerner, D. W., \& Sargent, A. I. 1995.  Submillimeter Continuum 
Observations of the T Tauri Spectroscopic Binary GW Orionis.  \aj, 
109:2655--2669.}

\ref{Mathieu, R. D., Martin, E. L., \& Maguzzu, A. 1996.  UZ Tau E: A New 
Classical T Tauri Spectroscopic Binary. 
\baas 188:60.05.} 

\ref{Mathieu, R. D., Stassun, K., Basri, G., Jensen, E. L. N., Johns-Krull, 
C. M., Valenti, J. A., \& Hartmann, L. W. 1997.  The Classical T Tauri 
Spectroscopic Binary DQ Tau. I. Orbital Elements and Light Curves.  \aj, 
113:1841--1854.} 

\ref{Meyer, M. R., \& Beckwith, S. V. W. 1998. The Environments of Pre-Main-Sequence
Stars:  Brown Dwarf Companions and Circumstellar Disk Evolution.
In {\refit Brown dwarfs and extrasolar planets, Proceedings of a Workshop held in
Puerto de la Cruz, Tenerife\/}, eds. R. Rebolo, E. L. Martin, Z. Osorio, M. Rosa, p.\ 245}

\ref{Milone, E.F. \& Mermilliod, J.--C. 1996. The origins, evolution, and 
destinies of binary stars in clusters. ASP Conference Series, 90.}

\ref{Moneti, A., \& Zinnecker, H. 1991, Infrared imaging photometry of binary 
T Tauri stars. \aap, 242:428--432.}

\ref{Monin, J.-L., M\'enard, F., \& Duch\^ene, G. 1998.  Using
  polarimetry to check rotation alignment in PMS binary stars. Principles of 
  the method and first results. \aap, 339:113--122.}

\ref{Neuh\"auser, R. 1997. Low-mass pre-main sequence stars and their X-ray 
emission. {\refit Science\/}, 276:1363-1370.}

\ref{Neuh\"auser, R. \& Brandner, W. 1998.  {\it HIPPARCOS\/} results for 
{\it ROSAT}-discovered young stars.  \aap, 330:L29--L32.} 

\ref{N\"urnberger, D., Brandner, W., Yorke, H. W., \& Zinnecker, H. 1998.  
Millimeter continuum observations of  X-ray selected T Tauri stars in 
Ophiuchus.  \aap, 330:549--558.} 

\ref{Osterloh, M. \& Beckwith, S. V. W. 1995.  Millimeter-wave continuum 
measurements of young stars.  \apj, 439:288--302.} 

\ref{Padgett, D. L., Strom, S. E., \& Ghez, A. 1997.  Hubble Space 
Telescope WFPC2 Observations of the Binary Fraction among Pre--Main-Sequence 
Cluster Stars in Orion.  \apj, 477:705--710.} 

\ref{Patience, J., Ghez, A. M., Reid, I. N., Weinberger, A. J., \& 
Matthews, K. 1998.  The Multiplicity of the Hyades and Its Implications for 
Binary Star Formation and Evolution.  \aj, 115:1972--1988.} 

\ref{Patience, J., Ghez, A. M., Reid, I. N., \& Matthews, K. 1998.  
Multiplicity Survey of Alpha Persei: Studying the Effects and Evolution of 
Companions.  \baas, 192:54.05.} 

\ref{Perrin, G., Coude du Foresto, V., Ridway, S. T., Mariotti, J.-M., Traub, W. A.,   
Carleton, N. P., \& Lacasse, M. G. 1998.  Extension of the effective temperature       
scale of giants to types later than M6.  \aap, 331:619-626}    

\ref{Petr, M. G., Coude Du Foresto, V., Beckwith, S. V. W., Richichi, A., 
\& McCaughrean, M. J. 1998.  Binary Stars in the Orion Trapezium Cluster 
Core.  \apj, 500:825--837.}

\ref{Popper, D. M. 1987.  A pre--main sequence star in the detached binary 
EK Cephei.  \apjl, 313:L81--L83.}

\ref{Prato, L. 1998. Pre--Main-Sequence Binaries and Evolution of their Disks.
  Ph.D. thesis, State Univ.\ of New York, Stony Brook.} 

\ref{Prato, L. \& Simon, M. 1997.  Are Both Stars in a Classic T Tauri 
Binary Classic T Tauri Stars?  \apj, 474:455--463.} 

\ref{Prosser, C. F., Stauffer, J. R., Hartmann, L., Soderblom, D. R., 
Jones, B. F., Werner, M. W., \& McCaughrean, M. J. 1994.  HST Photometry of 
the Trapezium Cluster.  \apj, 421:517--541.} 

\ref{Reipurth, B. \& Zinnecker, H. 1993.  Visual binaries among pre-main 
sequence stars.  \aap, 278:81--108.} 

\ref{Roddier, C., Roddier, F., Northcott, M. J., Graves, J. E., \& Jim, K. 
1996.  Adaptive Optics Imaging of GG Tauri: Optical Detection of the 
Circumbinary Ring.  \apj, 463:326--335.}

\ref{Rodriguez, L.F., D'Alessio, P.F., Wilner, D.J., Ho, P.T.P., Torrelles, J.M.,
Curiel, S., Gomez, Y., Lizano, S., Pedlar, A., Canto, J. \& Raga, A.C. 1998.
Compact protoplanetary disks in a binary system in L1551. {\refit
Nature\/}, 395:355-357.}

\ref{Schneider, G., Smith, B. A., Becklin, E. E., Koerner, D. W.,
 Meier, R., Hines, D. C., Lowrance, P. J., Terrile, R. J., Thompson,
 R. I., \& Rieke, M. 1999. NICMOS imaging of the HR 4796A
 circumstellar disk. \apjl, in press.}

\ref{Simon, M., Close, L.M., \& Beck, T.L. 1998. Adaptive optics imaging of
the Orion Trapezium cluster. \aj, in press.}

\ref{Simon, M., Ghez, A. M., \& Leinert, C. 1993.  Multiplicity and the 
ages of the stars in the Taurus star-forming region.  \apjl,
408:L33--L36.}

\ref{Simon, M., Ghez, A. M., Leinert, C., Cassar, L., Chen, W. P., Howell, 
R. R., Jameson, R. F., Matthews, K., Neugebauer, G., \& Richichi, A. 1995.  
A lunar occultation and direct imaging survey of multiplicity in the 
Ophiuchus and Taurus star-forming regions.  \apj, 443:625--637.} 

\ref{Simon, M., Holfeltz, S. T., \& Taff, L. G. 1996.  Measurement of T 
Tauri Binaries Using the Hubble Space Telescope Fine Guidance Sensors.  
\apj, 469:890--897.} 

\ref{Simon, M. \& Prato, L. 1995.  Disk Dissipation in Single and Binary 
Young Star Systems in Taurus.  \apj, 450:824--829.}

\ref{Stapelfeldt, K. R., Krist, J. E., Menard, F., Bouvier, J., Padgett, D. 
L., \& Burrows, C. J. 1998.  An Edge-On Circumstellar Disk in the Young 
Binary System HK Tauri.  \apjl, 502:L65--L69.} 

\ref{Swenson, F. J., Faulkner, J., Rogers, F. J., \& Iglesias, C. A. 1994.  
The Hyades lithium problem revisited.  \apj, 425:286--302.} 

\ref{Telesco, C. M., Fisher, R. S., Pina, R. K., Knacke, R. F., 
Dermott, S. F., Wyatt, M. C., Crogan, K., Ghez, A. M., Prato, L.,
Hartmann, L. W., \& Jayawardhana, R. 1998.  Mid-Infrared Imaging
with Keck II of the HR4796A Circumstellar Disk.  \apj, submitted.}

\ref{Terquem, C. \& Bertout, C. 1993.  Tidally-induced warps in T-Tauri 
disks - Part one - First order perturbation theory.  \aap, 274:291--303.}

\ref{Thiebaut, E., Balega, Y., Balega, I., Belkine, I., Bouvier, J., Foy, 
R., Blazit, A., \& Bonneau, D. 1995.  Orbital motion of DF Tauri from 
speckle interferometry.  \aap, 304:L17--L20.}

\ref{Walter, F. M., Vrba, F. J., Mathieu, R. D., Brown, A., \& Myers, P. C. 
1994.  X-ray sources in regions of star formation. 5: The low mass stars of 
the Upper Scorpius association.  \aj, 107:692--719.}

\ref{White, R. J., Ghez, A. M., Schultz, G., \& Reid, I. N. 1998.  
Spatially Resolved Spectroscopy of the PMS Quadruple GG Tau:  Evidence for 
a Substellar Companion.  \apj, in press.}

\ref{White, R. J. and Ghez, A. M. 1999. A Comparison of Circumprimary
and Circumsecondary Disks in Young Binary Systems. \baas, 193:73.11.}

\ref{Whitney, B. A. \& Hartmann, L. 1992.  Model scattering envelopes of
young stellar objects. I - Method and application to circumstellar disks.
\apj, 395:529--539.}

\ref{Wiegert, P. A. \& Holman, M. J. 1997.  The Stability of Planets in the 
Alpha Centauri System.  \aj, 113:1445--1450.} 

\bye